\documentclass[review,3p]{elsarticle}
\usepackage{blindtext}

\usepackage[T1]{fontenc}
\usepackage[utf8]{inputenc}
\usepackage[font=small]{caption}
\usepackage{verbatim}
\usepackage{float}
\usepackage{amsmath}
\usepackage{amsthm}
\usepackage{amssymb}
\usepackage{siunitx}
\usepackage{comment}
\usepackage{adjustbox}
\usepackage{mathdots}
\usepackage{graphicx}
\usepackage{fancyhdr}
\usepackage{amsfonts}
\usepackage{fixmath}
\usepackage{nicefrac}
\usepackage{ulem}
\usepackage{wasysym}

\usepackage{mdframed}
\newmdenv[
topline=false,
leftline=false,
rightline=false,
bottomline=false,
]{txtbox}

\usepackage{ragged2e}
\usepackage{booktabs}
\usepackage{array}
\usepackage{colortbl}

\usepackage{hyperref}
\hypersetup{colorlinks,citecolor=black,urlcolor=black,linkcolor=black} 
\newcommand{\ie}{\textit{i.e.}}

\newcommand{\cdp}{\textsc{CDP}}
\newcommand{\our}{\textsc{MInt}}
\newcommand{\meanshift}{\textsc{MeanShift}}
\newcommand{\wgl}{\textsc{WGL}}

\usepackage{multirow}

\usepackage{algorithm}
\usepackage{algpseudocode}
\algnewcommand\algorithmicforeach{\textbf{for each}}
\algdef{S}[FOR]{ForEach}[1]{\algorithmicforeach\ #1\ \algorithmicdo}

\usepackage{etoolbox}
\usepackage{tikz}
\usetikzlibrary{tikzmark}
\usetikzlibrary{calc}

\errorcontextlines\maxdimen

\newcommand{\ALGtikzmarkcolor}{black}
\newcommand{\ALGtikzmarkextraindent}{4pt}
\newcommand{\ALGtikzmarkverticaloffsetstart}{-.5ex}
\newcommand{\ALGtikzmarkverticaloffsetend}{-.5ex}
\makeatletter
\newcounter{ALG@tikzmark@tempcnta}

\newcommand\ALG@tikzmark@start{%
	\global\let\ALG@tikzmark@last\ALG@tikzmark@starttext%
	\expandafter\edef\csname ALG@tikzmark@\theALG@nested\endcsname{\theALG@tikzmark@tempcnta}%
	\tikzmark{ALG@tikzmark@start@\csname ALG@tikzmark@\theALG@nested\endcsname}%
	\addtocounter{ALG@tikzmark@tempcnta}{1}%
}

\def\ALG@tikzmark@starttext{start}
\newcommand\ALG@tikzmark@end{%
	\ifx\ALG@tikzmark@last\ALG@tikzmark@starttext
	\else
	\tikzmark{ALG@tikzmark@end@\csname ALG@tikzmark@\theALG@nested\endcsname}%
	\tikz[overlay,remember picture] \draw[\ALGtikzmarkcolor] let \p{S}=($(pic cs:ALG@tikzmark@start@\csname ALG@tikzmark@\theALG@nested\endcsname)+(\ALGtikzmarkextraindent,\ALGtikzmarkverticaloffsetstart)$), \p{E}=($(pic cs:ALG@tikzmark@end@\csname ALG@tikzmark@\theALG@nested\endcsname)+(\ALGtikzmarkextraindent,\ALGtikzmarkverticaloffsetend)$) in (\x{S},\y{S})--(\x{S},\y{E});%
	\fi
	\gdef\ALG@tikzmark@last{end}%
}

\apptocmd{\ALG@beginblock}{\ALG@tikzmark@start}{}{\errmessage{failed to patch}}
\pretocmd{\ALG@endblock}{\ALG@tikzmark@end}{}{\errmessage{failed to patch}}
\makeatother

\usepackage{resizegather}

\usepackage{xcolor}
\theoremstyle{definition}
\newtheorem{theorem}{Theorem}

\newtheorem{corollary}[theorem]{Corollary}
\newtheorem{proposition}[theorem]{Proposition}
\theoremstyle{definition}
\newtheorem{remark}{Remark}

\newtheorem{example}{Example}

\makeatother

\journal{Signal Processing}
\begin{document}
	\begin{frontmatter}
         \title{Labelled Multi-Object Fusion for Distributed Tracking in Partially Observable Heterogeneous Sensor Networks}
		
		\author[1]{Fei Chen\corref{cor2}}
		%\ead{fei.chen@adelaide.edu.au}
		\cortext[cor2]{First co-authors in alphabetical order and contributed equally to this work}
		
		\author[2]{Hoa Van Nguyen\corref{cor2}}
		%\ead{hoa.v.nguyen@curtin.edu.au}
		
		\author[3]{Alex S. Leong}
		%\ead{alex.leong@defence.gov.au}
		
		\author[3]{Sabita Panicker}
		%\ead{sabita.panicker1@defence.gov.au}
		
		\author[3]{Robin Baker}
		%\ead{robin.baker@defence.gov.au}
		
		\author[1]{Damith C. Ranasinghe\corref{cor1}}
		%\ead{damith.ranasinghe@adelaide.edu.au}
		\cortext[cor1]{Corresponding author}
		
		\address[1]{School of Computer Science and Information Technology, Adelaide University, Adelaide, SA 5005, Australia}
		\address[2]{School of Electrical Engineering, Computing, and Mathematical Sciences,
			Curtin University, Bentley, WA 6102, Australia}
		\address[3]{Defence Science and Technology Group, Australia}

		\begin{abstract}
			We consider the problem of using a distributed network of heterogeneous sensors with partial observability, limited communication bandwidth and computing power to track multiple, unknown, and time-varying numbers of objects. The fundamental problem in this context, distributed multi-object tracking (DMOT) to achieve \textit{globally consistent object states and labels or identities} at each sensor node, is non-trivial. We propose a new distributed framework for fusing labelled multi-object states in practical operating conditions. We formulate fusion as a multi-dimensional assignment problem. To overcome the inherent curse of dimensionality in the NP-hard problem, we devise a tractable hypothesis generation strategy exploiting geometric sparsity. Coupled with a novel cost structure to explicitly quantify cross-sensor association quality, this strategy translates the problem into an integer linear programming formulation, with a reduced solution space, solvable using standard algorithms. The rigorous formulation enables robust association across sensors without requiring prior knowledge of sensor topology, field of view information or full observability. Importantly, the proposed framework is agnostic to local tracking algorithms, allowing seamless integration with existing filtering methods while maintaining scalability. In contrast to current approaches seeking approximations via heuristics such as pairwise matching or clustering, we provide a theoretically grounded framework for the problem. Significantly, extensive experiments in representative tracking scenarios demonstrate that the proposed approach consistently improves the accuracy of fused trajectories and identity consistency compared to current state-of-the-art methods, whilst maintaining communication and computational tractability.
		\end{abstract} 
	\end{frontmatter}
	
	\section{Introduction} \label{sec:intro}
	Geographically distributed and heterogeneous sensor networks can realise persistent situational
    awareness in complex environments, critical for applications ranging from
    intelligent transport, maritime monitoring, environmental sensing and smart-city ecosystems
    to aerospace surveillance and defence~\cite{blackman1999design,AHLBERG200784,bar2011tracking}.
	In general, a distributed sensing capability overcomes challenges posed by fragmented and uncertain information due to the limited observation range of a single sensor or sensing modality, occlusions that lead to information loss, and dynamic operating conditions. Fundamental to situational awareness is a capability to track multiple, time-varying numbers of mobile objects manoeuvring under a network of sensors. Distributed tracking methods for multiple mobile objects, or distributed multi-object tracking (DMOT), aim to enable multiple sensor nodes to collaboratively estimate object trajectories from fragmented observations acquired across a network, in a manner that scales with the increasing number of sensor nodes and objects. By fusing complementary information from heterogeneous sensing modalities, DMOT improves trajectory continuity, robustness, and situational awareness beyond the capability of isolated sensors, while also distributing computational load across the network.
	
	Despite the advantages, \textit{the information fusion problem in DMOT is non-trivial}. Importantly, the fusion challenges extend beyond those encountered in conventional multi-object tracking (MOT), including object birth and death processes, false alarms, missed detections, and ambiguous
	measurement to object associations~\cite{mahler2007statistical}. Existing difficulties in MOT are amplified in decentralised sensing environments. Now, sensors operate independently, with limited and potentially unknown fields-of-view (FoVs), constrained communication and computing resources where consistent object identities are needed across fragmented and partially overlapping observations~\cite{he2018multi,yi2020distributed,chen2025distributed}. In practical sensor networks, local sensing nodes must independently estimate object states while simultaneously exchanging sufficient information to achieve network-wide consensus regarding object identities, trajectories, and kinematic states~\cite{li2022multisensor}. Consequently, a prevalent challenge in modern information fusion is constructing a globally consistent view of multiple dynamic objects at each sensor node from distributed, and often only partially overlapping, local observations, under resource constraints imposed by sensors. In particular, developing a practical solution to the  fusion problem requires addressing several important challenges:
	\begin{itemize}
		\item Each node observes only a partial and localised view of the environment; consequently, fusion across heterogeneous sensors with unknown FoVs is inherently hard. One strategy is to exchange and fuse raw measurements. However, the approaches are often sensitive to asynchronous updates and measurement ordering~\cite{linh2024}. Further, fusing does not distribute the computational load whilst requiring the sharing of sensor characteristics. An alternative is density fusion~\cite{uney2013dist,battistelli2013consensus}, which requires careful management of both shared and complementary FoV information to avoid information redundancy and the associated ``data incest'' problem~\cite{yi2020distributed,TAYLOR2019}.
		\item A distributed fusion framework must jointly achieve kinematic consistency---\textit{state}---and label consistency---\textit{identity}~\cite{hoa2021distributed, li2017clustering, dunham2004multiple} for the same physical object across multiple sensors to support reliable communication of object information and downstream decision-making.
		\item A fusion algorithm must remain effective under practical deployment constraints, including limited communication bandwidth, restricted computational resources and decentralised processing at sensor nodes~\cite{uney2013dist}.
	\end{itemize}
	
	The pertinent challenges imply that a distributed fusion approach needs to balance a difficult trade-off between statistical consistency and computational tractability. Recent fusion approaches are mostly based on the Random Finite Set (RFS) framework. RFS-based algorithms offer a theoretical generalisation of single-object distributed estimation to multi-object settings~\cite{mahler2000optimal}. Most distributed RFS fusion methods perform density fusion via Geometric Averaging (GA)~\cite{uney2013dist,battistelli2013consensus,CHAI2023101956} or Arithmetic Averaging (AA)~\cite{li2017clustering,Gao2020multiobject,LI2024102199}. While these approaches provide elegant probabilistic formulations, they incur significant communication and computational overheads, and their performance deteriorates in heterogeneous sensor networks with limited or partially overlapping FoVs~\cite{yi2020distributed,uney2019fusion}. These limitations have motivated the development of alternative, scalable, track-level fusion approaches.
	
	Track-to-track fusion methods improve scalability by exchanging compact track estimates rather than full posterior densities; however, this leads to a multi-dimensional assignment problem (MDAP) for multiple nodes. Consequently, optimal solutions are NP-hard; therefore, existing methods rely on heuristic approximations. Recent \textit{online} approaches, formulated under unknown FoV sensor settings, mitigate complexity by relying on sequential pairwise associations~\cite{hoa2021distributed} or clustering~\cite{chen2025distributed,ester1996density}. Pairwise matching results are selection-order dependent, whereas clustering approximates global associations using local similarity measures. As a result, clutter, missed detections, fragmented tracks and partially overlapping FoVs can degrade global association accuracy and lead to inconsistent label identities~\cite{he2018multi,chen2025distributed}.
	
	In this paper, we propose a novel track-to-track fusion framework by formulating distributed multi-object tracking under limited, heterogeneous, and unknown sensor fields of view (FoVs) as a multi-dimensional assignment problem.
	We exploit the geometric sparsity inherent in limited FoV conditions to formulate a sparse MDAP, enabling an efficient integer linear programming solution that jointly resolves associations across all sensors without resorting to heuristic clustering or order-dependent pairwise fusion strategies used in prior approaches. The main contributions from this study are summarised as follows:
	
	\begin{itemize}
		\item[\textit{i)}] We formulate \textit{online} labelled track fusion as a sparse multi-dimensional assignment problem, efficiently solvable using integer linear programming. 
		\item[\textit{ii)}] We introduce a cost formulation, relating fusion costs to estimation errors, enabling computationally tractable global optimisation and principled gating parameter selection. Further, we propose a tractable hypothesis-generation strategy based on geometric sparsity and clique discovery to prune the MDAP search space prior to optimisation, thereby enabling efficient global association across multiple sensors.
		\item[\textit{iii)}] The proposed sparse MDAP formulation simultaneously resolves global multi-sensor associations, eliminating the order dependence of sequential pairwise fusion while reducing communication overhead by exchanging track estimates rather than full multi-object densities.
		\item[\textit{iv)}] We demonstrate that the proposed approach consistently improves the accuracy of fused trajectories and identity consistency compared to current state-of-the-art methods, whilst maintaining communication and computational tractability, through a rigorous experimental regime that includes simulated and real-world trajectory data from taxis and drones in 3D tracking scenarios.
	\end{itemize}
	
	The paper is structured as follows. Section~\ref{sec:background} reviews related work in multi-object tracking, assignment problems, as well as distributed tracking. Section~\ref{sec:proposed_method} details our proposed method for labelled multi-object fusion via integer linear programming.  Section~\ref{sec:experiment} presents detailed experimental results and discussion. Section~\ref{sec:conclusion} concludes our findings and discusses potential future directions.
	
	\section{Background and Related Work} \label{sec:background}
	This section reviews the background most relevant to our work: multi-object tracking (MOT), the multi-dimensional assignment problem in MOT, and distributed multi-object tracking (DMOT) methods.
	
	\subsection{MOT Frameworks}
    For tracking multiple objects \textit{at a single sensor node}, various techniques have emerged. These include Global Nearest Neighbour (GNN)~\cite{blackman1999design}, Belief Propagation (BP)~\cite{meyer2018message}, Joint Probabilistic Data Association (JPDA)~\cite{bar2009probabilistic}, Multiple Hypothesis Tracking (MHT)~\cite{blackman1999design,reid1979algorithm}, and Random Finite Set (RFS) based methods~\cite{mahler2007statistical}. Traditional methods, such as JPDA and MHT, must determine measurement-to-object associations before utilising single-object filters. In particular, JPDA employs a probabilistic model to compute association likelihoods but lacks robust mechanisms to handle object births and deaths. MHT considers multiple association hypotheses over time~\cite{reid1979algorithm}; two notable variants, Hypothesis-Oriented MHT (HOMHT) and Track-Oriented MHT (TOMHT), are reviewed in~\cite{chong2019forty}.  
	Random finite set-based approaches represent multi-object states as random sets, circumventing explicit associations. Prominent RFS-based filters include Probability Hypothesis Density (PHD)~\cite{vo2006gaussian}, its cardinalised extension (CPHD), the Multi-Bernoulli (MB)~\cite{mahler2007statistical} and Poisson Multi-Bernoulli Mixture (PMBM) families~\cite{garcia2018poisson}, Labelled Multi-Bernoulli (LMB)~\cite{reuter2014lmb}, and Generalized Labelled Multi-Bernoulli (GLMB) filters~\cite{vo2013glmb}, the last providing the first exact closed-form solution to the labelled multi-object filtering recursion (see~\cite{vo2024overview} for a detailed treatment). 
	
	Concurrently, the computer vision community has made substantial progress in detection-based multi-object tracking through efficient tracking-by-detection frameworks. Representative methods, such as SORT~\cite{bewley2016simple} and its successors, improve data association through motion prediction, appearance embeddings, cascaded matching, and joint detection--embedding architectures. While these methods achieve impressive accuracy and real-time performance in centralised visual tracking, they rely on a single sensing platform and therefore do not address the challenges of distributed multi-sensor track fusion across heterogeneous, limited FoVs.
	
	Notably, the fusion strategy we investigate is \textit{agnostic} to the tracking approach at each sensor node. Therefore, the formulation we pursue can co-exist with multiple different MOT methods, even within the same sensor network.
	
	\subsection{Multi-Dimensional Assignment Problem (MDAP)}
    A core challenge in multi-object tracking (MOT) is the data association problem, namely determining the optimal assignment between observations and existing object tracks~\cite{blackman1999design}. When formulated over multiple scans or multiple sensors, this assignment problem is commonly cast as a Multi-Dimensional Assignment Problem (MDAP), which is NP-hard in general~\cite{poore1994multidimensional}. Despite its computational complexity, \textit{centralised} MOT has demonstrated that practically tractable solutions can be obtained using Integer Linear Programming (ILP) and related optimisation frameworks. Rather than eliminating the inherent NP-hardness of the problem, ILP formulations exploit the structure and sparsity of practical data association problems---introduced, for example, through gating and problem-specific constraints---allowing modern optimisation algorithms to efficiently solve instances of practical size~\cite{wolsey2014integer}. Consequently, ILP and its variants, including linear programming relaxations, network-flow mixed integer programming and Lagrangian relaxation for minimum-cost network flow~\cite{wolsey2014integer}, have become established approaches for data association problems in \textit{centralised} MOT. 
	
	However, these formulations cannot be directly extended to \textit{distributed} MOT (DMOT) because the underlying association problem is fundamentally different. In centralised MOT, the objective is to associate \textit{sensor measurements with object tracks}~\cite{blackman1999design}; in contrast, in distributed MOT, the objective is to associate local state estimates produced by multiple sensing nodes~\cite{chong1990distributed}. Consequently, the association cost is no longer defined by the likelihood of assigning measurements to tracks, but by the consistency between local tracking outputs. Moreover, the associated uncertainty is driven by estimation errors and sensor heterogeneity, leading to fundamentally different constraints on associations. Although principled gating remains an effective mechanism for reducing the association search space in both settings, existing ILP-based approaches are designed for measurement-to-track association and therefore do not naturally accommodate estimate-to-estimate association across sensors in distributed tracking. Our study develops a problem formulation and a tractable solution for the DMOT fusion problem.
	
	\subsection{DMOT Frameworks}
    Modern DMOT methodologies, frequently RFS-based, pursue density fusion through Geometric Averaging (GA)~\cite{uney2013dist,battistelli2013consensus,GAO2023} or Arithmetic Averaging (AA)~\cite{li2017clustering,Gao2020multiobject,li2021}. These techniques, bearing resemblance to Fr\'{e}chet means~\cite{li2022multisensor}, face considerable hurdles with elevated computational and communication demands~\cite{chen2025distributed,hoa2021distributed} and falter when confronted with dissimilar FoVs---GA methods preserve solely common objects~\cite{Gao2020multiobject,uney2019fusion}, whilst AA methods suffer degradation in the absence of FoV alignment~\cite{yi2020distributed}.
    Notably, fusing \textit{labelled} multi-object densities, irrespective of whether sensor nodes share a FoV, for the comparison of labelled densities from different sensor nodes. But labels are discrete random variables defined by users or sensor nodes to uniquely identify multiple objects, and they can vary across sensor nodes. Standard divergences (e.g., Kullback–Leibler divergence) are undefined for labelled densities with different supports and fusing labelled densities whilst maintaining global identity consensus is not trivial.
	
	Motivated by these limitations, recent research explores track-to-track fusion, in which compact local track estimates are exchanged rather than full multi-object densities, thereby substantially reducing communication and computational overhead.
	Historically, track-to-track fusion relied on the restrictive assumption that no false tracks or missed objects occur, thereby requiring an identical number of local tracks across any two nodes~\cite{chong1990distributed}. Relaxing these assumptions naturally leads to a multi-dimensional assignment problem, for which globally optimal solutions are computationally intractable in general. Consequently, contemporary methods rely on heuristic approximations, such as pairwise matching~\cite{kaplan2008assignment} or  clustering~\cite{he2018multi,ester1996density}. Yet, these heuristic approximations can compromise global association consistency and struggle to maintain stable object labels across distributed sensors. 
	Although Nguyen et al~\cite{nguyen2026themean} developed a DMOT algorithm for fusing multi-object trajectories via the concept of Fr\'{e}chet mean, the approach is not suitable for online estimation and operates under the condition of known sensor FoVs. 

    Inevitably, fusing tracks or labelled multi-object estimates requires addressing the label-consistency problem---the challenge of arriving at a set of globally unique identities across a sensor network. Recent studies propose achieving label consistency and post-kinematic consistency, in which label information is not considered, through weighted graph labelling~\cite{chen2025distributed}, while still relying on heuristic assumptions about the density of cluster centres~\cite{rodriguez2014clustering}.

    In contrast to existing methods, our approach formulates track-level fusion as a sparse global assignment problem, jointly resolving multi-sensor associations without heuristic clustering or sequential pairwise fusion, whilst naturally accommodating heterogeneous sensors with limited and unknown FoVs. Since our work focuses on developing a new formulation for the DMOT problem, we employ our proposed fusion approach with weighted-graph labelling to achieve label consistency across a network of sensors.

    \subsection{Notations} 
	We briefly introduce notational background here, before delving into the proposed method. In this article, a lowercase symbol (e.g., $x$) denotes a single-object state, while an uppercase symbol (e.g., $X$) denotes a multi-object state.  Bold italic symbols (e.g., $\boldsymbol{x}, \boldsymbol{X}$) represent labelled states, whereas other bold symbols (e.g., $\mathbf{H,~b,y}$) represent vectors or matrices. Blackboard symbols (e.g., $\mathbb{X}, \mathbb{L}$) denote the underlying spaces. 
    
	\section{Distributed Multi-Object Track Fusion} \label{sec:proposed_method}
	
	This section describes the proposed distributed multi-object track fusion framework, which aims to combine local track estimates from multiple sensor nodes into a globally consistent, single-integrated view of the environment across a network of sensors. Rather than relying on heuristic clustering or sequential pairwise fusion, the proposed method formulates cross-sensor association as a tractable Multi-Dimensional Assignment Problem (MDAP) and solves it using Integer Linear Programming (ILP). 

    	\begin{figure}[!htbp]
		\centering \includegraphics[width=0.95\textwidth]{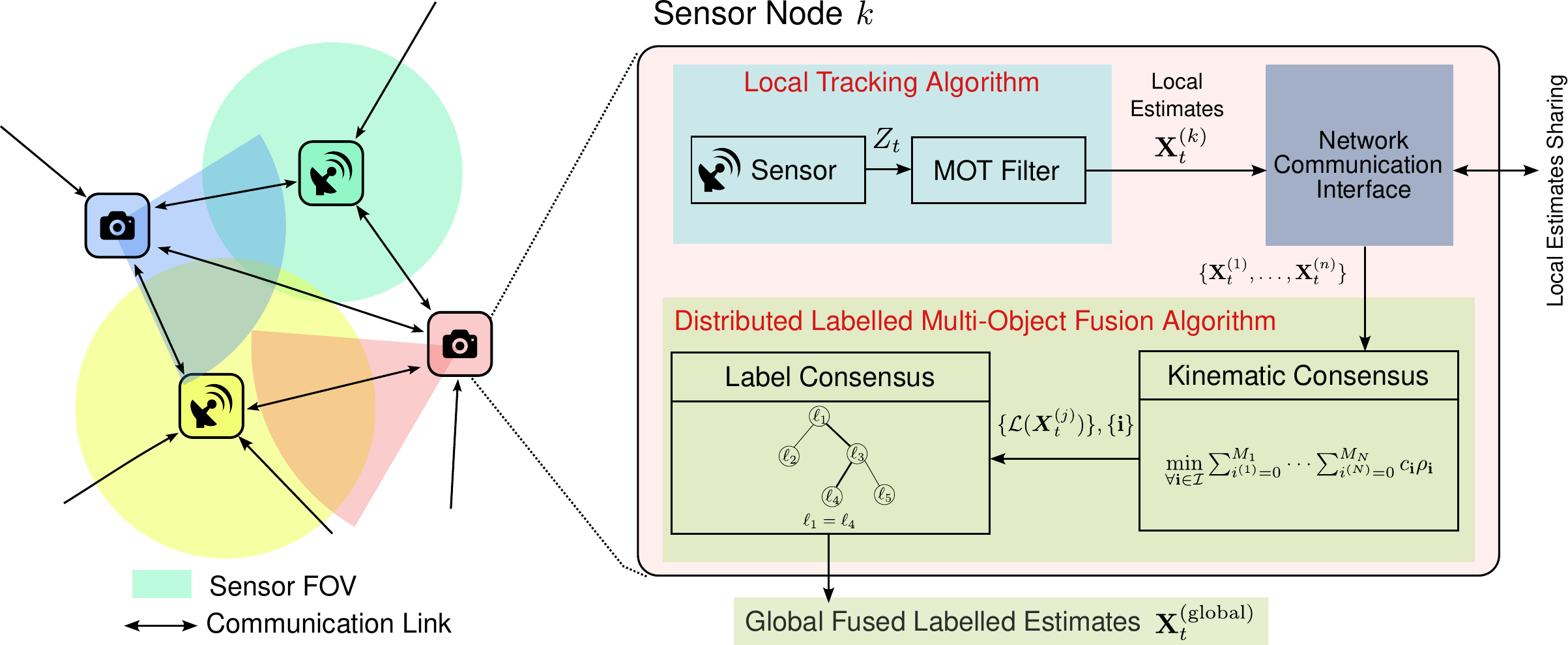}
		\caption{A distributed sensor network system with limited FoVs sharing local track estimates with each node implementing the proposed distributed fusion algorithm.}
	\label{fig:network_model} 
\end{figure}

    In the proposed framework for a distributed fusion algorithm illustrated in Figure~\ref{fig:network_model}, each sensor node produces a set of local track estimates using its local tracking algorithm or simply a tracker. Then, spatial gating is used to construct a compatibility graph across sensors, from which candidate cross-sensor association hypotheses are generated via clique enumeration and subsequently pruned to maintain solution tractability. The remaining hypotheses are assembled to formulate the global association problem as a sparse MDAP solved as an ILP to obtain globally consistent associations. These associations determine the fused object states through \textit{kinematic consensus} and establish consistent track labels across the network through \textit{label consensus}. We elaborate on our approach next.

	\subsection{System Model} 
	Consider a Wireless Sensor Network (WSN) as depicted in Figure~\ref{fig:network_model}. The WSN is represented by an undirected graph, $G_{\text{net}} = (\mathcal{N},\mathcal{A})$, where $\mathcal{N}$ is the set of network nodes and $\mathcal{A} \subset \mathcal{N}\times\mathcal{N}$ comprises the arcs symbolising connections among these nodes. This network performs surveillance, observing and tracking an unknown, fluctuating set of mobile objects across a large area. Each node is equipped with a sensor having a restricted FoV, which may also produce false alarms and misdetections. Furthermore, each node incorporates a dedicated processing unit that computes local multi-object state estimates. It is essential to note that no central fusion entity exists in this network; instead, each node is equipped with a transceiver, enabling direct or indirect exchange of multi-object state estimates via standard \textit{ad-hoc} or mesh networking protocols.

At discrete time instant $t$, each node $n \in \mathcal{N}$ calculates its local labelled multi-object state estimate, denoted as $\boldsymbol{X}^{(n)}_t \subset \mathbb{X} \times \mathbb{L}$. Here, $\mathbb{X}$ is the object state space (e.g., position, velocity), and $\mathbb{L}$ is the label space. For notational conciseness, the temporal index $t$ is omitted when clarity is maintained. We denote $M_n = |\boldsymbol{X}^{(n)}|$ and express $\boldsymbol{X}^{(n)} = \{ \boldsymbol{x}^{(n)}_1,\dots, \boldsymbol{x}^{(n)}_{M_n}\}$.

Each individual local, labelled single-object state is represented by $\boldsymbol{x}^{(n)} = (x^{(n)},\ell^{(n)})$, where $x^{(n)} \in \mathbb{X}$ denotes the non-labelled (unlabelled) single-object state, and $\ell^{(n)} = (b,\beta,n) \in \mathbb{L}$ serves as the object label. Here, $b$ is a positive integer denoting the birth time, and $\beta$, a non-negative integer, distinguishes among multiple objects appearing simultaneously at time $b$. The set of all such local labels is given by $\mathcal{L}(\boldsymbol{X}^{(n)}) = \{ \ell^{(n)}_1,\dots, \ell^{(n)}_{M_n}\}$, where $\mathcal{L}: \mathbb{X} \times \mathbb{L} \to \mathbb{L}$ is the label projection function defined by $\mathcal{L}((x,\ell))=\ell$, and $\mathcal{L}(\boldsymbol{X}) = \bigcup_{\boldsymbol{x} \in \boldsymbol{X}} \mathcal{L}(\boldsymbol{x})$.
Additionally, the set of all \textit{unlabelled} multi-object states (i.e., the state components without labels) is defined as $\mathcal{S}(\boldsymbol{X}^{(n)}) = \{x_1^{(n)},\dots,x_{M_n}^{(n)}\}$, where $\mathcal{S}: \mathbb{X} \times \mathbb{L} \to \mathbb{X}$ is the state projection function given by $\mathcal{S}((x,\ell)) = x$, and $\mathcal{S}(\boldsymbol{X}) = \bigcup_{\boldsymbol{x}\in \boldsymbol{X}}\mathcal{S}(\boldsymbol{x})$.

\subsection{Multi-Dimensional Assignment Problem (MDAP)}
Our objective is to fuse the local estimates $\{ \mathcal{S}(\boldsymbol{X}^{(n)}) \}_{n \in \mathcal{N}}$, originating from $N=|\mathcal{N}|$ nodes, to establish a single integrated view~\cite{dunham2004multiple} representing the global multi-object state, $\boldsymbol{\bar{X}}$.

To ground the problem in a theoretical framework, we consider casting DMOT fusion as a general set-partitioning problem, following the framework in~\cite{poore2000multidimensional} for MOT. Interestingly, the resulting set-partitioning problem of seeking the optimal disjoint subsets can be formulated as a multi-dimensional assignment problem by accounting for the practical reality, especially in \textit{partially} observable settings, that not all sensor nodes will estimate all objects. Crucially, in the present problem setting, the limited-field-of-view sensors induce strong \textit{spatial sparsity} in the association structure, since only geographically compatible track estimates can correspond to the same physical object. Then, exploiting the inherent sparsity allows us to formulate a cost structure and a method to generate a comprehensive set of feasible association hypotheses without an explosion in the search space for solutions to the assignment problem. Consequently, the effective dimensionality of the MDAP is substantially reduced, and only a limited set of feasible association variables needs to be considered. Subsequently, we can reformulate the resulting MDAP as a \textit{sparse} Integer Linear Programming (ILP) problem, in which the objective is to select an optimal subset of hypotheses that yields a globally consistent assignment of all local sensor estimates. 
For such sparse block-structured formulations, efficient solution methods are available under appropriate structural conditions~\cite{koutecky2018parameterized}. By rendering the assignment problem practically tractable, our formulation allows us to jointly resolve associations across all sensors, rather than through local grouping decisions or order-dependent pairwise matching.

Formally, we seek an optimal partition of the cumulative set of local estimates $\{ \mathcal{S}(\boldsymbol{X}^{(n)}) \}_{n \in \mathcal{N}}$ into disjoint subsets, encoded by $\mathcal{I}^* = \{ \mathbf{i}_1, \dots, \mathbf{i}_{|\mathcal{I}^*|} \}$ where $|\mathcal{I}^*|$ is unknown. Each $\mathbf{i} \in \mathcal{I}^*$ is an $N$-tuple of indices $\mathbf{i} = (i^{(1)}, \dots, i^{(N)})$ that associates estimates from different nodes to a common physical object. To formalise the search for such a partition as an $N$-dimensional optimisation problem, let $\mathbb{I} \triangleq \{0,1,\dots,M_1\} \times \dotsm \times \{0,1,\dots,M_N\}$ denote the space of all possible association vectors, so that $\mathcal{I}^* \subseteq \mathbb{I}$. For any candidate selection $\mathcal{I} \subseteq \mathbb{I}$, we introduce a binary association variable $\rho_{\mathbf{i}}$ for every $\mathbf{i} \in \mathbb{I}$, indicating whether $\mathbf{i}$ is selected:
\begin{align}
	\rho_{\mathbf{i}} = 
	\begin{cases} 
		1 & \text{if } \mathbf{i} \in \mathcal{I} \\ 
		0 & \text{otherwise.} 
	\end{cases}
\end{align}

Conversely, $\mathcal{I} = \{\mathbf{i} \in \mathbb{I} \mid \rho_{\mathbf{i}} = 1\}$, therefore the candidate partitions and indicator vectors $\{\rho_{\mathbf{i}}\}_{\mathbf{i} \in \mathbb{I}}$ are in one-to-one correspondence. Consequently, finding the optimal partition $\mathcal{I}^{*}$ is equivalent to solving for the optimal indicators $\{\rho^{*}_{\mathbf{i}}\}_{\mathbf{i} \in \mathbb{I}}$ and the resulting problem can be expressed as the following Multi-Dimensional Assignment Problem (MDAP):

\begin{align} \label{eq:map_main}
    \{\rho^{*}_{\mathbf{i}}\} = \underset{\{\rho_{\mathbf{i}}\}}{\arg\min} ~ \sum_{\mathbf{i} \in \mathbb{I}} c_\mathbf{i} \rho_{\mathbf{i}}
\end{align}
subject to
	\begin{align}
		\sum_{\mathbf{i} \in \mathbb{I} \,:\, i^{(n)} = j}
		\rho_{\mathbf{i}} &= 1, \quad \forall\, n \in \mathcal{N},\ \forall\, j \in \{1,\dots,M_n\}, \label{eq:map_card}\\
		\rho_{\mathbf{i}} &\in \{0,1\}, \quad \forall\, \mathbf{i} \in \mathbb{I}, \label{eq:map_binary}
\end{align}
where $c_{\mathbf{i}}$ is the cost of association $\mathbf{i}$.  Recall, each $ N$-tuple index in $\mathbf{i}$ has a fixed length $N$ with each component representing an estimate of a unique object from a node. A critical challenge in distributed sensing is that an object is rarely detected by all nodes simultaneously due to limited FoVs or occlusions~\cite{hoa2021distributed}. Consequently, we need to explicitly accommodate the physical reality in the MDAP formulation and adopt the concept of dummy reports~\cite{poore2000multidimensional}.

\noindent\textbf{Dummy Reports.~}
To account for the physical reality of partial observability while maintaining a rigid $N$-dimensional assignment structure, we adapt the concept of a \textit{dummy estimate} (or dummy report), denoted by the index $0$ for each node $n$ (i.e., $x^{(n)}_0$).  

The dummy estimate serves as a mathematical placeholder representing ``missing data'' or ``non-detection'' for a specific node within a hypothesis. This allows us to construct valid $N$-tuples even when only a subset of nodes detects the object. Crucially, the constraints in \eqref{eq:map_card} enforce that every \textit{actual} local estimate ($i^{(n)} > 0$) is assigned to exactly one global hypothesis (partitioning the real data). However, no such cardinality constraint is imposed on the dummy reports ($i^{(n)} = 0$). They act as a source of infinite capacity, allowing them to be reused arbitrarily across multiple hypotheses to account for independent missed objects across the network. 

Then, to solve \eqref{eq:map_main}, an \textit{effective} association cost $c_{\mathbf{i}} \in \mathbb{R}$ and the corresponding global estimate $\bar{x}_{\mathbf{i}}$ must be defined for each possible association vector $\mathbf{i}$. 

\noindent\textbf{Association Cost and Global Estimates.~}
For any association vector $\mathbf{i}$, let $\mathcal{M}_{\mathbf{i}} = \{ n \in \mathcal{N} \mid i^{(n)} > 0 \}$ denote the subset of nodes contributing actual estimates and $d(\cdot,\cdot)$ denote any norm-induced metric, taken as the Euclidean distance throughout our experiments. 

As the number of possible assignments scales exponentially with the number of sensors $N$, the problem becomes computationally unmanageable. However, we recognise that the problem space afforded by the limited sensor FoVs in practical sensing scenarios is naturally sparse. Consequently, we can effectively employ gating to prune unlikely associations, retaining only those indicators $\rho_{\mathbf{i}}$ whose contributing estimates are mutually separated by no more than a distance bound $\gamma_{\max}$, i.e.

\begin{align}\label{eq:gating_cond}
	\max_{n,m \in \mathcal{M}_{\mathbf{i}}} d(x^{(n)}_{i^{(n)}}, x^{(m)}_{i^{(m)}}) \le \gamma_{\max}.
\end{align}

For the retained associations, we define $c_{\mathbf{i}}$ and $\bar{x}_{\mathbf{i}}$ based on $\mathcal{M}_{\mathbf{i}}$ as follows:
\begin{itemize}
	\item \textbf{Null Assignment} ($|\mathcal{M}_{\mathbf{i}}| = 0$): For cases where $i^{(n)} = 0 ~\forall n \in \mathcal{N}$, the global estimate is defined as $\emptyset$ and the association cost is set to the gating bound $\gamma_{\max}$ to preclude selection during minimisation. Hence:
	\begin{align} \label{eq:x_glob_case1}
		\bar{x}_\mathbf{i} = \emptyset; \quad c_\mathbf{i} = \gamma_{\max}.
	\end{align}
	Null Assignment is included for completeness of the fixed $N$-dimensional formulation.
	\item \textbf{Singleton Assignment} ($|\mathcal{M}_{\mathbf{i}}| = 1$): This corresponds to cases where exactly one $i^{(m)} > 0$ for some $m \in \mathcal{N}$, and $i^{(n)} = 0$ for all $n \neq m$. This hypothesis represents a valid object located within the exclusive FoV of a single node $m$, remaining unobserved by all other nodes. Let $x^{(m)}_{i^{(m)}}$ be this unique estimate, then the global estimate retains this sole estimate, and the association cost is set to the maximum gating cost $c_\mathbf{i} = \gamma_{\max}$, penalising ungrouped estimates:
	\begin{align} \label{eq:x_glob_case2}
		\bar{x}_\mathbf{i} = x^{(m)}_{i^{(m)}}; \quad c_\mathbf{i} = \gamma_{\max}.
	\end{align}
	Since \eqref{eq:gating_cond} ensures $c_\mathbf{i} \le \gamma_{\max}$ for every retained multi-lateral hypothesis, the cost allocation allows favouring multi-node consensus wherever the gate admits it.
	\item \textbf{Multi-Lateral Assignment} ($|\mathcal{M}_{\mathbf{i}}| \ge 2$): The global estimate $\bar{x}_{\mathbf{i}}$ is defined as the centroid of the associated estimates, and the association cost $c_{\mathbf{i}}$ is derived from the mean deviation of these estimates from their centroid, i.e.
	\begin{align} \label{eq:x_glob_case3}
		\bar{x}_{\mathbf{i}} = \frac{1}{|\mathcal{M}_{\mathbf{i}}|} \sum_{n \in \mathcal{M}_{\mathbf{i}}} x^{(n)}_{i^{(n)}}; \quad  c_{\mathbf{i}} = \frac{1}{|\mathcal{M}_{\mathbf{i}}|} \sum_{n \in \mathcal{M}_{\mathbf{i}}} d(x^{(n)}_{i^{(n)}}, \bar{x}_{\mathbf{i}}).
	\end{align}
\end{itemize}

\noindent In practice, determining an appropriate gating bound $\gamma_{\max}$ is necessary. We rely on Proposition~\ref{prop:1}, which relates the gating bound to the estimation error of the network of sensors and the fused estimates.

\begin{proposition}[Estimation Error and Gating] \label{prop:1}
	\textit{Equip the object state space $\mathbb{X}$ with a norm $\|\cdot\|$, whose linear structure permits the centroid below to be formed, and let $d(a, b) := \|a - b\|$ be the induced metric. } Let $y \in \mathbb{X}$ be the true state of an object. For a sensor network with node set $\mathcal{N}$, let $X = \{x^{(n)} \mid n \in \mathcal{M}\}$ be the estimated states of object $y$ produced by a non-empty subset of nodes $\mathcal{M} \subseteq \mathcal{N}$, where $x^{(n)} \in \mathbb{X}$ is node $n$'s estimate of $y$. Suppose there exists a constant $e_{\max} > 0$ such that the maximum estimation error for object $y$ over the estimates in $X$ is bounded with probability at least $\alpha \in [0, 1]$, i.e.,
		\begin{align*}
			\textrm{Pr}\left(\max_{x \in X} d(x^{}, y) \le e_{\max}\right) \ge \alpha.
		\end{align*}
		Then, defining the centroid of the estimates in $X$ as $\bar{x} = \frac{1}{|X|} \sum_{x \in X} x$, and the mean deviation from the centroid as $ e_X = \frac{1}{|X|} \sum_{x \in X} d(x, \bar{x})$, the following probabilistic bound holds:
		\begin{align*}
			\textrm{Pr}(e_X \le 2e_{\max}) \ge \alpha.
	\end{align*}
\end{proposition}

\begin{proof}
	Let $A$ be the event that $\max_{x \in X} d(x, y) \le e_{\max}$, so that
		$\textrm{Pr}(A) \ge \alpha$, and suppose $A$ occurs. Since the metric is induced by a norm,
		\[
		d(y, \bar{x})
		= \Big\| \tfrac{1}{|X|} \sum_{x \in X} (y - x) \Big\|
		\le \frac{1}{|X|} \sum_{x \in X} \| y - x \|
		\le e_{\max}.
		\]
		Hence, for every $x \in X$, the triangle inequality gives
		$d(x, \bar{x}) \le d(x, y) + d(y, \bar{x}) \le 2 e_{\max}$, and averaging over $X$ yields
		$e_X \le 2 e_{\max}$. Thus $A$ implies $e_X \le 2 e_{\max}$, and monotonicity of probability
		gives $\textrm{Pr}(e_X \le 2 e_{\max}) \ge \textrm{Pr}(A) \ge \alpha$.
\end{proof}

\begin{corollary}[Gate Admissibility]
	\label{cor:gate}
	Consider the setting in Proposition~\ref{prop:1} and let $A$ denote the event defined therein.  Then:
		\begin{equation}
			\Pr \Big( \max_{x_a, x_b \in X} d(x_a, x_b) \le 2 e_{\max} \Big) \ge \alpha .
		\end{equation}
\end{corollary}

\begin{proof}
	If event $A$ occurs, then $d(x, y) \le e_{\max}$ for every $x \in X$. Hence, for any $x_a, x_b \in X$, the triangle inequality gives
		\begin{equation}
			d(x_a, x_b) \le d(x_a, y) + d(y, x_b) \le 2 e_{\max} ,
		\end{equation}
		and taking the maximum over $x_a, x_b \in X$ shows that $A$ implies $\max_{x_a, x_b \in X} d(x_a, x_b) \le 2 e_{\max}$. Monotonicity of probability then yields the result, as in the proof of Proposition~\ref{prop:1}.
\end{proof}

As a consequence of Corollary~\ref{cor:gate}, with the gating threshold set to $\gamma_{\max} = 2 e_{\max}$, the local estimates of a common object $y$ are mutually admissible under the gating condition~\eqref{eq:gating_cond}, and the corresponding association is therefore not pruned by gating, with probability at least $\alpha$. Whether it is subsequently retained as a candidate hypothesis depends on the tractable hypothesis generation strategy in Section~\ref{sec:hyp_gen}.

\begin{remark}
	Corollary~\ref{cor:gate} rests on the triangle inequality alone, whereas Proposition~\ref{prop:1} additionally requires the vector-space operations that define the centroid, together with the convexity of $z \mapsto d(y,z)$, both supplied by the norm. The threshold $\gamma_{\max} = 2e_{\max}$ therefore applies to the Euclidean and Mahalanobis distances under a common weighting matrix, but not to divergences, such as the Kullback--Leibler divergence, which are not norm-induced and need not satisfy either symmetry or the triangle inequality. The derived gating threshold provides a useful default for pruning implausible associations without heuristic parameter tuning.
\end{remark}

\begin{remark} \label{rem:alpha} 
    The choice of the parameter $\alpha$ is informed by the statistical distribution of the estimation error. Suppose each node's error is zero-mean Gaussian with covariance $\sigma^{2} I_{D}$, where $D$ is the state dimension and $\sigma$ is the standard deviation of each state component. The normalised squared error $d^{2}(x^{(n)}, y)/\sigma^{2}$ then follows a chi-squared distribution ($\chi^{2}_{D}$) with $D$ degrees of freedom, since it is a sum of $D$ independent squared standard normal variables. Setting $e_{\max} = k\sigma$ therefore bounds the error at a single node with probability $p = \Pr(\chi^{2}_{D} \le k^{2})$, which increases with $k$ and decreases with $D$. For instance, at $k = 3$, $p \approx 98.9\%$ for $D = 2$ and $97.1\%$ for $D = 3$, whereas at $k = 5$ the corresponding values are $99.9996\%$ and $99.9985\%$, respectively. Since the bound must hold simultaneously at all $|X|$ nodes contributing estimates of the object, the union bound gives $\alpha \ge 1 - |X|(1 - p)$; consequently, we can observe $k$ between $3$ and $5$ to serve as a reasonable default, with $\sigma$ taken as the largest local standard deviation across the contributing nodes. The quality of the bound therefore depends on the performance of a local tracker---the accuracy of the local covariance estimates.
\end{remark}

\begin{figure}[h!]
	\centering
	\includegraphics[width=1\textwidth]{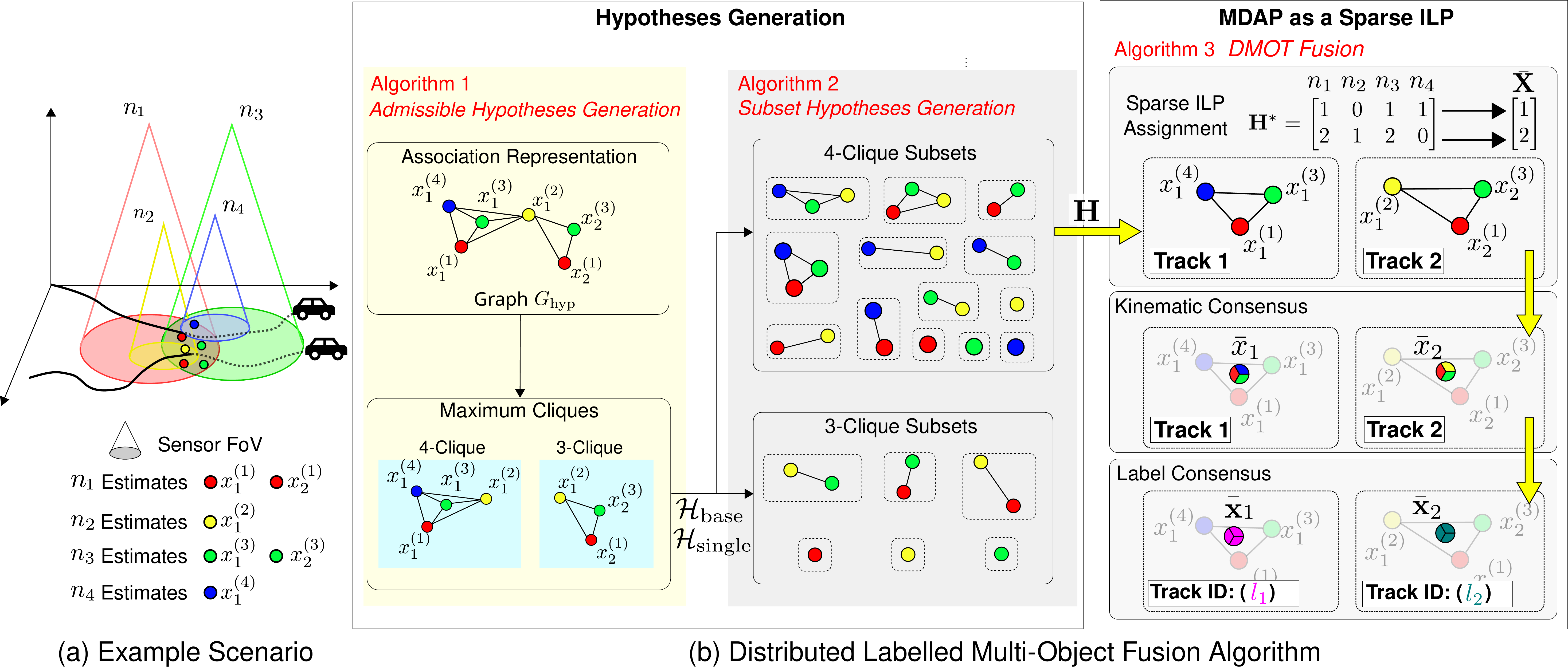}
	\caption{\textbf{a)} Example scenario in which four sensor nodes $n_1, \dots, n_4$ with limited FoV track two objects. 
    \textbf{b)}~Overview of the proposed fusion algorithm. \textit{(1)~Construct an association representation.~}Shared and local estimates form the vertices of graph $G_{\text{hyp}}$ where an edge joins estimates from different nodes satisfying the gating condition in~\eqref{eq:gating_cond}.
    Consequently, admissible hypotheses are its cliques. \textit{(2)~Construct candidate hypotheses.~}First, we extract the maximal cliques (a $4$-clique and a $3$-clique in our example) to form the base hypotheses $\mathcal{H}_{\text{base}}$ and singletons $\mathcal{H}_{\text{single}}$.   
    Subsequently, we refine the clique-based hypothesis by enumerating their $k_\text{best}$ subsets. The retained candidate subset $\mathbf{H}$ are used to select a least-cost partition $\mathbf{H}^{*}$ using the ILP formulation in Section~\ref{sec:mdap_ilp}. 
    The selected estimates are fused to achieve kinematic ($\bar{x}_1$ and $\bar{x}_2$ in the example) and label consensus ($\bar{\rm{\mathbf{x}}}_1$ and $\bar{\rm{\mathbf{x}}}_2$).}
	\label{fig:alg_example}
\end{figure}

. 

Although we have arrived at an MDAP formulation for the distributed multi-object tracking problem, solving the problem directly as formulated in~\eqref{eq:map_main}--\eqref{eq:map_binary} is computationally infeasible for non-trivial scenarios. Because, it involves one variable $\rho_{\mathbf{i}}$ for each of the $|\mathbb{I}| = \prod_{n \in \mathcal{N}}(M_n + 1)$ association vectors. For example, considering the San Francisco taxi-tracking scenario in Section~\ref{sec:exp2}, with $N = 40$ nodes and only $20$ local estimates per sensor node, this amounts to roughly $10^{53}$ variables. This exponential growth in $N$ and $M_n$ is known as the \textit{curse of dimensionality}. We discuss our proposed approach to remedy the pertinent challenge in the context of DMOT, next.

\subsection{Feasible Hypothesis Generation} \label{sec:hyp_gen}
We introduce a computationally tractable method that generates a compact set of feasible association hypotheses \textit{prior} to solving the assignment problem. Effectively, the approach reduces the solution space by discarding the vast majority of \textit{geometrically} invalid associations, allowing the subsequent optimisation to operate on a much smaller set of candidates using a graph-based formulation. Figure~\ref{fig:alg_example} illustrates the approach using four limited-FoV sensors observing two mobile objects. Overall, our approach solves the problem in two key sequential stages:
\begin{itemize}
    \item[(i)]  \textit{Construct an Association Representation.~}We encode pairwise associations of the multi-object estimates---including shared and local estimates at a node---as an undirected graph, 
    $G_{\text{hyp}}$, to facilitate generating the possible set of association hypotheses. We consider the gating bound described in~\eqref{eq:gating_cond} to determine edges, since this determination is pairwise, a graph representation loses no information. 
    \item[(ii)] \textit{Construct Candidate Hypotheses.~}Our graph representation captures association hypotheses based on spatio-temporal alignment. We employ the graph structure to generate candidate hypotheses by eliminating those that are \textit{redundant} (shared estimates across overlapping cliques) and prioritising effort to consider estimates with the largest evidence of association to an object, in two stages. The first extracts a compact cover of the admissible hypotheses and the second refines it, as that cover can be ambiguous, by enumerating over the subsets of admissible hypotheses.
\end{itemize} 

\begin{algorithm}[!h]
	{\footnotesize{}{}{}
		\caption{AdmissibleHypothesesGeneration}
		\label{alg:base_hyp}
		\begin{algorithmic}[1]
			\State \textbf{Input}: $\{\mathcal{S}(\boldsymbol{X}^{(n)})\}_{n \in \mathcal{N}}$, $\gamma_{\max}$ \Comment{local estimates of all nodes, gating bound}
			\State \textbf{Output}: Base hypotheses $\mathcal{H}_{\text{base}}$, singleton hypotheses $\mathcal{H}_{\text{single}}$ \Comment{collections of subsets of $V$}
            \State $V \gets \{(n,j) : n \in \mathcal{N},\, j \in \{1,\dots,M_n\}\}$, with $x_{(n,j)} := x^{(n)}_j$ \Comment{vertices carry node of origin}
			\State $E \gets \emptyset$
			\ForAll{pairs $v = (n,j),\, w = (m,k) \in V$ where $n < m$}
			\If{$d(x_v, x_w) \le \gamma_{\max}$}
			\State $E \gets E \cup \{(v,w)\}$
			\EndIf
			\EndFor
			\State $G_{\text{hyp}} \gets (V, E)$
			\State $\mathcal{H}_{\text{base}} \gets \Call{BronKerbosch}{G_{\text{hyp}}}$ \Comment{find all maximal cliques}
			\State $\mathcal{H}_{\text{single}} \gets \{\{v\} \mid v \in V\}$ \Comment{construct singleton hypotheses}
			\State \Return $\mathcal{H}_{\text{base}}$, $\mathcal{H}_{\text{single}}$
	\end{algorithmic}}
\end{algorithm}

We describe the construction of an association representation in Section~\ref{sec:graph} and Algorithm~\ref{alg:base_hyp}, and the approach for constructing the candidate hypothesis in Section~\ref{sec:clique} and Algorithm~\ref{alg:base_hyp2}.

\subsubsection{Graph-Based Formulation for Association Representations} \label{sec:graph}
Importantly, conditions defining an admissible hypothesis are pairwise: i)~the gate in~\eqref{eq:gating_cond} constrains pairs of estimates; 
ii)~each node contributes at most one estimate to a hypothesis, which rules out pairs from the same node. A hypothesis is therefore admissible precisely when all of its pairs are compatible, so recording compatibility once per pair loses no information. 
A graph structure can conveniently record exactly the necessary information, with estimates as vertices and compatible pairs as edges. Consequently, enumerating admissible hypotheses simply becomes a clique enumeration problem. Since limited FoVs render most pairs incompatible, the graph is sparse, making clique enumeration efficient in practice.

Accordingly, we construct an undirected graph $G_{\text{hyp}} = (V,E)$, as illustrated in Figure~\ref{fig:alg_example} and described in Algorithm~\ref{alg:base_hyp}. A vertex $v = (n,j) \in V$ denotes the $j$th estimate of node $n$, with state $x_v \triangleq x^{(n)}_j$. An edge joins $v = (n,j)$ and $w = (m,k)$ if and only if $n \neq m$ and $d(x_v,x_w) \le \gamma_{\max}$, so the two estimates are admissible as observations of a common object. A vertex set $\mathcal{C} \subseteq V$ is a valid hypothesis if no two of its vertices share a node index and $\max_{v,w \in \mathcal{C}} d(x_v,x_w) \le \gamma_{\max}$. These two conditions hold exactly when every pair in $\mathcal{C}$ is adjacent, so an interesting facet of the graph structure is that admissible hypotheses are precisely the cliques of $G_{\text{hyp}}$ with at most one estimate per node.

\subsubsection{Clique-Based Candidate Hypothesis Generation}\label{sec:clique}

A maximal clique is one that cannot be extended by any adjacent vertex. Since every clique lies within a maximal one, the maximal cliques implicitly represent the admissible hypotheses. Therefore, we first consider maximal cliques as possible estimate-to-object associations. We extract the maximal cliques that cover every admissible hypothesis, along with all singletons needed to capture objects seen by only one node. However, maximal cliques alone are insufficient. For example, an estimate gated to two objects can belong to several cliques as seen by $x_1^{(2)}$ belonging to both \textit{4-Clique} and \textit{3-Clique} in Figure~\ref{fig:alg_example}, but the correct hypothesis must be a strict subset of one. Therefore, in the next stage, we enumerate subsets, retaining the $k_{\text{best}}$ largest since their number grows exponentially. 
Importantly, as we illustrate in Figure~\ref{fig:alg_example}, retaining subsets allows the partitioning stage to reach the correct assignment. For instance, $x^{(2)}_{1}$, which belonged to both \textit{4-Clique} and \textit{3-Clique}, is assigned to Track 2 after the sparse ILP assignment optimisation, so Track 1 is recovered as a three-element subset of the \textit{4-Clique}. 

Our hypothesis generation, following a systematic, sequential refinement approach in two stages, is illustrated in Figure~\ref{fig:alg_example}. The first stage detailed in Algorithm~\ref{alg:base_hyp} generates base hypotheses $\mathcal{H}_{\text{base}}$ by identifying all maximal cliques using the Bron--Kerbosch algorithm~\cite{bron1973algorithm}.\footnote{While clique-finding is generally NP-hard, the limited sensor FoVs and strict gating threshold $\gamma_{\max}$ render the compatibility graph $G_{\text{hyp}}$ highly sparse, making the application of Bron--Kerbosch computationally efficient in practice.} It also returns every estimate as a singleton hypothesis in $\mathcal{H}_{\text{single}}$, which is crucial for modelling objects detected by a single node, even in the presence of limited FoV or occlusion, and which guarantees that~\eqref{eq:ilp} admits a feasible partition. The second stage, described in Algorithm~\ref{alg:base_hyp2}, resolves this ambiguity. Since~\eqref{eq:map_card} permits a shared estimate in only one selected hypothesis, the hypothesis for the other object is a strict subset of its maximal clique, so subsets must be considered as feasible candidate hypotheses.

\begin{algorithm}[!tb]
	{\footnotesize{}
		\caption{HypothesesSubsetGeneration}
		\label{alg:base_hyp2}
		\begin{algorithmic}[1]
			\State \textbf{Input}: Base hypotheses $\mathcal{H}_{\text{base}}$, singleton hypotheses $\mathcal{H}_{\text{single}}$, budget $k_{\text{best}}$
			\State \textbf{Output}: Candidate hypothesis matrix $\mathbf{H}$
			\State $\mathcal{P}\leftarrow\emptyset$ \Comment{pool of candidate subsets}
			\ForAll{$\mathcal{C}\in\mathcal{H}_{\text{base}}$}
			\State $\mathcal{H}_{\text{sub}} \leftarrow$ first $k_{\text{best}}$ subsets $\mathcal{C}' \subseteq \mathcal{C}$ with $|\mathcal{C}'| \ge 2$, by descending $|\mathcal{C}'|$\Comment{terminate at budget}
			\State $\mathcal{P}\leftarrow\mathcal{P}\cup\mathcal{H}_{\text{sub}}$ \Comment{union discards shared subsets}
			\EndFor
			\State $\mathcal{P}\leftarrow\mathcal{P}\cup\mathcal{H}_{\text{single}}$ \Comment{singletons guarantee feasibility of \eqref{eq:ilp}}
			\State Enumerate $\mathcal{P} = \{\mathcal{C}_1,\dots,\mathcal{C}_K\}$; $\mathbf{H} \gets [\mathbf{h}_1;\dots;\mathbf{h}_K]$ with $h^{(n)}_k = j$ if $(n,j) \in \mathcal{C}_k$, else $0$
			\State \Return $\mathbf{H}$
	\end{algorithmic}}
\end{algorithm}

Enumerating all subsets is infeasible, a clique of size $|\mathcal{C}|$ has $2^{|\mathcal{C}|}$ subsets. Therefore, we seek to prioritise efforts to consider estimates with the largest evidence of association to an object by retaining only the $k_{\text{best}}$ by adapting $k$-best hypothesis management in classical MHT~\cite{bar2011tracking}. We consider prioritising by enumerating subsets in descending cardinality. This ordering follows from our cost structure. By~\eqref{eq:x_glob_case2}--\eqref{eq:x_glob_case3}, a hypothesis covering $m$ gated estimates costs at most $\gamma_{\max}$ but the same estimates left as singletons cost $m\gamma_{\max}$. Consequently, the savings from grouping grow with subset size. However, naively increasing $k_{\text{best}}$ is counterproductive; it adds hypotheses of near-identical cost, worsening association ambiguity and track switching, and increasing the size of the MDAP. We examine this trade-off in Section~\ref{sec:param_study}.  

The output of Algorithm~\ref{alg:base_hyp2} is a $K\times N$ candidate hypothesis matrix $\mathbf{H}$,
\begin{align} \label{eq:H_matrix}
	\mathbf{H} = \begin{bmatrix}
		h^{(1)}_{1} & \dots  & h^{(N)}_{1}\\
		\vdots & \ddots & \vdots\\
		h^{(1)}_{K} & \dots  & h^{(N)}_{K}
	\end{bmatrix} =
	\begin{bmatrix}
		\mathbf{h}_1\\
		\vdots \\
		\mathbf{h}_K
	\end{bmatrix},
\end{align}
where each row $\mathbf{h}_k = [h^{(1)}_k, \dots, h^{(N)}_k]$ encodes one candidate subset, $K$ is the number of candidate hypotheses retained, and $h^{(n)}_k \in \{0, 1, \dots, M_n\}$ is the index of node $n$'s local estimate in hypothesis $k$, with $0$ denoting a dummy estimate.

\subsection{DMOT Multi-Dimensional Assignment as a Sparse Integer Linear Programming Problem}~\label{sec:mdap_ilp}
Owing to the sparsity induced by the hypothesis generation stage, only a limited set of feasible association variables needs to be considered. Thus, substantially reduces the effective dimensionality of the MDAP. With the tractable set of $K$ candidate association hypotheses generated and compiled into the matrix $\mathbf{H}$, the original MDAP in \eqref{eq:map_main} can be reformulated as a sparse Integer Linear Programming (ILP) problem where the objective is to select an optimal subset of hypotheses from $\mathbf{H}$ that yields a globally consistent assignment of all local estimates. In practice, this optimisation is readily solved by a standard branch-and-bound algorithm, and the selected hypotheses determine the globally consistent associations between local track estimates, from which the fused object states are computed.

Let $\mathbf{C} = [c_{i_1},\dots,c_{i_K}]^{\top}$ be the cost vector (defined in  \eqref{eq:x_glob_case1}--\eqref{eq:x_glob_case3}) corresponding to each hypothesis row $\mathbf{h}_k$ in $\mathbf{H}$. The optimisation problem is:
\begin{align} \label{eq:ilp}
	\min_{\mathbf{y} \in \{0,1\}^{K} } \mathbf{C}^{\top} \mathbf{y} \text{ subject to } \mathbf{A} \mathbf{y} = \mathbf{b},
\end{align}
where $\mathbf{y} = [y_1,\dots,y_K]^{\top}$ is the binary selection vector indicating whether to select ($y_k=1$) or discard ($y_k=0$) the $k$-th hypothesis from $\mathbf{H}$. The constraint system ensures that every \textit{actual} local estimate is assigned to exactly one selected hypothesis:
\begin{itemize}
	\item $\mathbf{b} = [1,\dots,1]^{\top}$ is the all-ones vector of length $M_{\text{tot}}$, where $M_{\text{tot}} = \sum_{n \in \mathcal{N}} M_n$ is the total number of actual local estimates, so that $M_{\text{tot}} = |V|$.
	\item $\mathbf{A} \in \{0,1\}^{M_{\text{tot}} \times K}$ is the binary constraint matrix, whose rows are indexed by the vertices $v \in V$ of $G_{\text{hyp}}$: $A_{v,k} = 1$ if $v \in \mathcal{C}_k$, that is, if the local estimate $x_v$ is included in the $k$-th candidate hypothesis, and $A_{v,k} = 0$ otherwise.
\end{itemize}

Because Algorithm~\ref{alg:base_hyp2} retains every singleton hypothesis, every vertex $v \in V$ belongs to at least one $\mathcal{C}_k$, so every row of $\mathbf{A}$ has a non-zero entry. Selecting all $M_{\text{tot}}$ singleton rows therefore satisfies $\mathbf{A}\mathbf{y} = \mathbf{b}$, so \eqref{eq:ilp} is always feasible and an optimal solution exists. If $\mathbf{y}^*$ is the optimal solution to \eqref{eq:ilp}, then the solution to the MDAP is the set of association vectors corresponding to the non-zero elements of $\mathbf{y}^*$:
\begin{align} \label{eq:map_sol}
	\mathbf{H}^* = \{ \mathbf{h}_k \in \mathbf{H} \mid y^*_k = 1 \}.
\end{align}

\begin{algorithm}[!tb]
	{\footnotesize{}{}\caption{DMOT Fusion via ILP}
		\label{algo:dmot_fusion}
		\begin{algorithmic}[1]
			\State \textbf{Input}: $\{ \boldsymbol{X}^{(n)} \}_{n \in \mathcal{N}}, G_{\text{wgl}}$ \Comment{local estimates of all nodes, weighted graph label}
			\State \textbf{Output}: $\boldsymbol{\bar{X}}, G_{\text{wgl}}$ \Comment{global labelled multi-object estimates, weighted graph label}
			\vspace{0.1cm}
			\State $\mathbf{H}^* \leftarrow \Call{SolveILP}{\{\mathcal{S}(\boldsymbol{X}^{(n)})\}_{n\in\mathcal{N}}}$; \Comment{$\mathbf{H}$ via Algorithms.~\ref{alg:base_hyp}--\ref{alg:base_hyp2}; solve \eqref{eq:ilp}}
			\State $\bar{X} \gets\{ \bar{x}_k| \mathbf{h}_k \in \mathbf{H}^* \} $ via \eqref{eq:x_glob_case1}--\eqref{eq:x_glob_case3} \Comment{compute global kinematic estimates}
			\State $\{\boldsymbol{L}^{(n)}\}_{n \in \mathcal{N}} \gets \{ \mathcal{L}(\boldsymbol{X^{(n)}}) \}_{n \in \mathcal{N}}$ \Comment{extract local labels }
			\State $(G_{\text{wgl}}, \boldsymbol{\bar{L}}) \gets \textsc{WGL}(G_{\text{wgl}}, \mathbf{H}^*, \{\boldsymbol{L}^{(n)}\}_{n \in \mathcal{N}})$; \Comment{update graph and resolve global labels via \wgl}
			\State $\boldsymbol{\bar{X}} \gets\{ (\bar{x}_k, \bar{\ell}_k) \mid \mathbf{h}_k \in \mathbf{H}^* \}.$ \Comment{construct final global labelled states}
	\end{algorithmic}}
\end{algorithm}

\subsection{Global Labelled State Reconstruction}
The final stage reconstructs the global Multi-Object State $\boldsymbol{\bar{X}}$ from the optimal associations $\mathbf{H}^*$. We perform reconstruction in two steps: kinematic consensus and label consensus. Detailed steps are depicted in Algorithm~\ref {algo:dmot_fusion}. 

\noindent\textbf{Kinematic Estimates.~}
The cardinality of the global set is determined by the number of selected hypotheses, i.e., $|\bar{X}| = \|\mathbf{y}^*\|_0$. For each selected hypothesis $\mathbf{h}_k \in \mathbf{H}^*$, the global kinematic state $\bar{x}_k$ is computed as the centroid of its constituent local estimates, as defined in \eqref{eq:x_glob_case1}--\eqref{eq:x_glob_case3} and illustrated in Figure~\ref{fig:alg_example}(b).

\noindent\textbf{Label Consensus.~}
To ensure a single integrated view, we must resolve label ambiguities among the distributed nodes. For each global track $\bar{x}_k$, we aggregate the set of associated local labels $\boldsymbol{L}_k = \{ \ell^{(n)} \mid h_k^{(n)} > 0 \}$. To determine a unique global label $\bar{\ell}_k$ that minimises label switching, we employ the Weighted Graph Labelling (\wgl) strategy~\cite{chen2025distributed}. This method maintains a historical record of associations via an undirected weighted graph $G_{\text{wgl}}=(V_{\text{wgl}}, E_{\text{wgl}}, w)$, where vertices $V_{\text{wgl}}$ represent unique local labels and edges $E_{\text{wgl}}$ signify historical co-occurrence in fused hypotheses. Upon forming a global kinematic cluster, the graph is updated by reinforcing consistent associations: if an edge exists between two labels in the cluster, its weight is decremented ($w \leftarrow \max(0, w-1)$) to reflect higher affinity; otherwise, a new edge is initialised with a maximum penalty $w_{\max}$. Global consensus is then achieved by identifying the connected component corresponding to the cluster's smallest local label and assigning the component's lexicographically minimum label as the unique global identifier $\bar{\ell}_k$, ensuring persistent identification across the network history. The output of the method is the set of labelled global states:
\begin{align}
	\boldsymbol{\bar{X}} = \{ (\bar{x}_k, \bar{\ell}_k) \mid \mathbf{h}_k \in \mathbf{H}^* \}.
\end{align}

\section{Experiments} \label{sec:experiment}
This section evaluates the proposed distributed multi-object tracking (DMOT) fusion framework using a series of simulated experiments in both two-dimensional (2D) and three-dimensional (3D) environments. The experiments are designed to assess the robustness of the proposed MDAP-based fusion method under varying sensing conditions and to demonstrate its effectiveness in resolving association ambiguities across distributed sensors. Without loss of generality, for all experiments, the labelled multi-Bernoulli (LMB) filter is employed as the local tracker at each sensor node. The LMB filter provides labelled track estimates that are shared and fused using our proposed approach.

\vspace{2mm}
\noindent
\textbf{Environment Settings.~} 
Each run is performed on an Intel(R) Core(R) Ultra 9 285K CPU and MATLAB-2024b. Within the LMB filter, single-object filtering is performed using the standard Kalman filter (KF)~\cite{kalman1960} when the object dynamic model and measurement model are both linear (e.g., position sensors), and the unscented Kalman filter (UKF)~\cite{julier1997new} when either model is non-linear (e.g., radar sensors). The sparse ILP assignment problem in~\eqref{eq:ilp} is solved using MATLAB's \texttt{intlinprog} branch-and-bound solver. All fusion algorithms are implemented natively in MATLAB, so the reported fusion times are directly comparable.

Since the initial object's locations are unknown, we apply the adaptive birth procedure (ABP) for the LMB filter~\cite{reuter2014lmb}. Notably, we focus on implementing these filters for \textit{real-time} applications; thus, filtering parameters are chosen to ensure the processing time of each filter is smaller than the measurement interval ($T_0=1$~s in this case). Below are our fine-tuned\footnote{LMB filters are fine-tuned with the utmost effort to ensure a fair comparison of tracking accuracy and processing time for practical applications.} filtering parameters, which enable the LMB filter to run in real time and limit the maximum number of hypotheses to below $1000$. For adaptive birth, the expected number of birth objects at each time is $0.5$ while the maximum existence probability of a newborn object is set at $0.01$. 

\vspace{0.2cm}

\noindent
\textbf{Performance Evaluation and Comparison Methods.~}We compare the proposed algorithm, we refer to as 

\underline{M}ulti-object fusion with \underline{M}ultidimensional assignment using Sparse Linear \underline{Int}eger Programming (\our) with Weighted Graph Labelling (\our-\wgl), with \cdp-\wgl{} and \meanshift-\wgl{} in~\cite{chen2025distributed}. Our selection of baselines follows the evaluation in~\cite{chen2025distributed}, which proposed \cdp-\wgl{} and provided comparative results for MeanShift-\wgl{} and found both methods to outperform the DBSCAN-based approach of~\cite{he2018multi} and the sequential pairwise fusion in~\cite{hoa2021distributed}. Notably, pairwise fusion is disadvantaged on structural grounds, since its outcome depends on the order in which nodes are paired, which provides no consistent point of comparison for a method that resolves all sensors jointly. Consequently, we employ \cdp-\wgl{} and \meanshift-\wgl{} as current state-of-the-art DMOT algorithms for \textit{labelled} trackers under limited FoV sensors. Both of these approaches use cluster analysis to mitigate the computational complexity of the MDAP. 

Density-based RFS fusion methods are likewise unsuitable, as they require full multi-object densities and correspondingly higher communication bandwidth than the low-bandwidth distributed setting considered here. All three algorithms employ the same \wgl{} label consensus stage~\cite{chen2025distributed} and differ only in how cross-sensor associations are formed, so differences in OSPA\textsuperscript{(2)} are attributable to the association step rather than to the labelling strategy.
Additionally, to appreciate the performance of the DMOT method, as an ideal reference, we also consider a centralised tracking algorithm, MS-LMB~\cite{vo-msglmb2019}, with complete knowledge of the sensor characteristics.

We evaluate the tracking performance via two metrics:  i) Optimal Sub-Pattern Assignment (OSPA)~\cite{schuhmacher2008consistent}, and ii) OSPA-on-OSPA (OSPA\textsuperscript{(2)})~\cite{beard2020a} with cut-off $c=100$~m, parameter $p=1$ and window length $10$~s. In particular, OSPA considers not only localisation errors but also cardinality (\ie, the number of objects) errors, whereas OSPA\textsuperscript{(2)} accounts for localisation, cardinality, track fragmentation, and track label-switching errors. A smaller metric value indicates a better tracking performance. Additionally, we report the average fusion time at each measurement time step for the fusion algorithms considered. Notably, reported results are averaged over $100$ Monte Carlo (MC) runs. Next, we report the results from an extensive series of experiments conducted to:
\begin{itemize}
    \item Understand the failure modes of computationally efficient heuristic-based fusion approaches for DMOT (Section~\ref{sec:exp-fail-modes}).
    \item Empirically validate and demonstrate the practical utility of Proposition~\ref{prop:1}  (Section~\ref{sec:prop1_validation}).
    \item Understand the impact of $k_{\text{best}}$ subset selection  (Section~\ref{sec:param_study}).
    \item Evaluate and compare performance in 2D DMOT scenarios using CRAWDAD (San Francisco, U.S.A.) and Porto (Porto, Portugal) taxi datasets  (Section~\ref{sec:exp2}).
    \item Evaluate and compare performance in a 3D DMOT scenario tracking drones with a network of heterogeneous sensors  (Section~\ref{sec:exp-drones}).
\end{itemize}

\subsection{Failure Modes of Clustering-Based Fusion}\label{sec:exp-fail-modes}
\begin{figure}[!htb]
	\centering \includegraphics[width=1.0\textwidth]{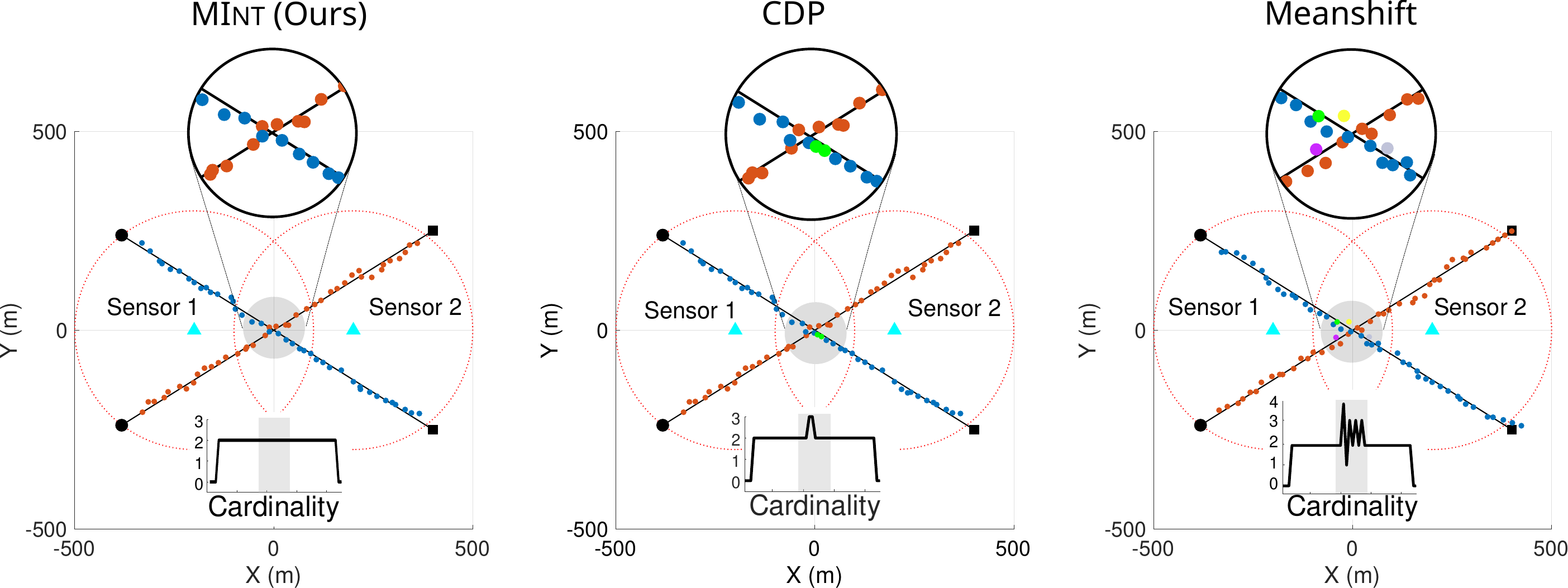}
	\caption{Two objects on intersecting trajectories observed by two FoV-limited sensors. Insets show the estimates near the crossing point and their corresponding cardinalities. \cdp{} leaves one estimate unassociated, producing a spurious track (green). MeanShift leads to large cardinality fluctuations. Our formulation, \our, consistently maintains the correct association and cardinality.}
	\label{fig:failure_example} 
\end{figure}

To illustrate the limitations of clustering-based distributed multi-object tracking approaches, we consider the object-crossing scenario in Figure~\ref{fig:failure_example}. In this setup, two objects move along intersecting trajectories within the sensing regions of two sensors. The configuration provides a challenging case for evaluating each algorithm's ability to handle the fusion problem at closely spaced objects at the crossing point.

We can observe that the \cdp{} method fails to correctly fuse the object estimates at the crossing point. One estimate is not properly associated during fusion, resulting in an additional false track, indicated by the green dots, and an incorrect cardinality estimate. This behaviour is attributable to the use of the k-means clustering procedure within \cdp{}, which forms the basis for inferring both the number and the locations of cluster centres. During close-proximity interactions, k-means incorrectly infers the cluster structure, leading to improper partitioning of track estimates and subsequent fusion errors.

The MeanShift-based method exhibits more pronounced degradation at the crossing, with larger fluctuations in the cardinality estimate. This sensitivity can be attributed to the bandwidth selection in the MeanShift algorithm. When objects are spatially close, a fixed bandwidth may either merge distinct clusters or fragment a single cluster, thereby introducing fusion errors and degrading cardinality consistency.

In contrast, the proposed approach maintains correct data association throughout the crossing event and preserves an accurate cardinality estimation. This behaviour suggests that the global assignment formulation provides improved robustness in scenarios where local clustering decisions become ambiguous. Overall, the example demonstrates the effectiveness of the proposed framework in handling close-proximity multi-object interactions.

\subsection{Proposition Validation}\label{sec:prop1_validation}
\begin{figure}[!tbh]
	\centering
	\includegraphics[width=1.0\linewidth]{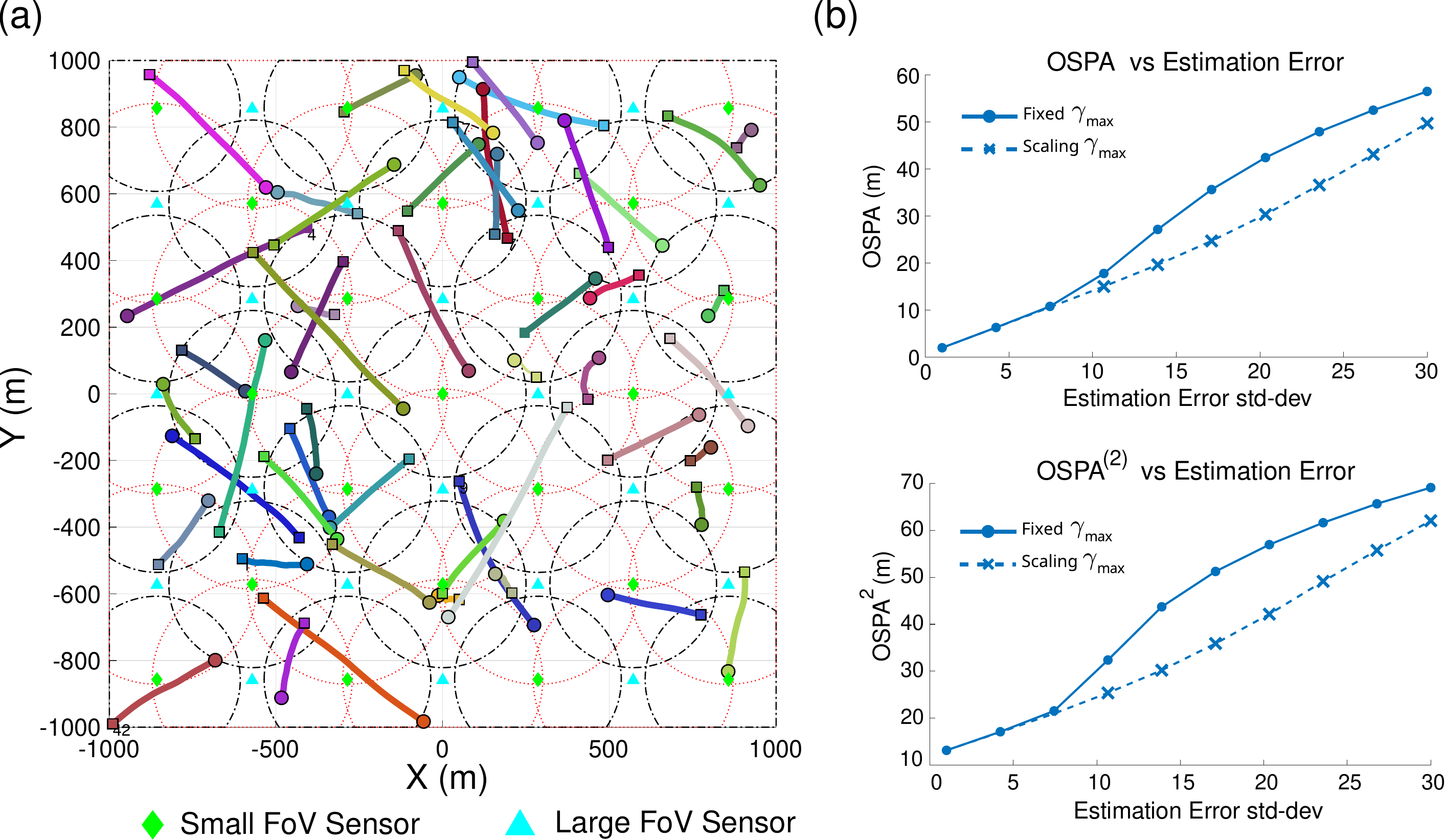}
	\caption{Proposition validation. (a)~Experiment scenario; (b)~Comparison of OSPA and OSPA\textsuperscript{(2)} between fixed and scaling $\gamma_{\max}$ according to measurement noise (sensor estimation accuracy) }
	\label{fig:proposition_validation}
\end{figure}

We investigate Proposition~\ref{prop:1} to demonstrate its practical utility using a synthetic scenario with the ability to control the pertinent estimation error settings as illustrated in Figure~\ref{fig:proposition_validation}(a). In this experiment, $25$ sensor nodes with a maximum detection range of $\SI{300}{\meter}$ and $24$ sensor nodes with a maximum detection range of $\SI{250}{\meter}$ are uniformly distributed over a $\SI{2}{\kilo\meter} \times \SI{2}{\kilo\meter}$ area. A total of $50$ objects evolve according to a constant-velocity dynamic model.

To enable precise control and analysis of the estimation errors at each sensor node, we do not employ local LMB filters for tracking. Instead, each node's state estimate is modelled as an additive white Gaussian perturbation of the ground truth, \ie{} $x_{\text{est}} = x_{\text{truth}} + w$, where $w\sim\mathcal{N}(0_{2\times1}, \sigma_{x}^{2}I_{2\times2})$. The maximum estimation error, denoted by $e_{\max}$, is defined as $3\sigma_x$, which corresponds to $\alpha > 98.9\%$ for $D = 2$ across all nodes following Remark~\ref{rem:alpha}, and thus provides a conservative upper bound on the estimation uncertainty.

Figure~\ref{fig:proposition_validation}(b) presents the mean OSPA and OSPA\textsuperscript{(2)} metrics of the fused tracks for two cases: i)~a fixed $\gamma_{\max}$; and ii)~a scaled $\gamma_{\max}$ that adapts to increasing estimation uncertainty. For the fixed setting, $\gamma_{\max}=\SI{6}{\meter}$ is used for all estimation error values and for the scaling $\gamma_{\max}$ setting, $\gamma_{\max}=2e_{\max}=6\sigma_{x}$ is used. As the estimation error increases, both approaches exhibit the expected performance degradation. However, the proposed scaling strategy consistently outperforms the fixed-parameter baseline across both metrics. These results support the practical usefulness of the bound in Proposition~\ref{prop:1} and demonstrate that adapting $\gamma_{\max}$ to the uncertainty provides a principled and effective mechanism for improving fusion performance.

\subsection{Parameter Studies} \label{sec:param_study}
The proposed algorithm includes a hyperparameter, $k_{\text{best}}$, to affect the number of feasible hypotheses generated during fusion. To assess the sensitivity of the algorithm to $k_{\text{best}}$, we conducted simulation experiments under the same conditions as detailed in Section~\ref{sec:prop1_validation}, using scenarios with $10$, $30$, and $50$ mobile objects and varying $k_{\text{best}}$ from $1$ to $10$.

\begin{figure}[!tbh]
	\centering
	\includegraphics[width=1.0\linewidth]{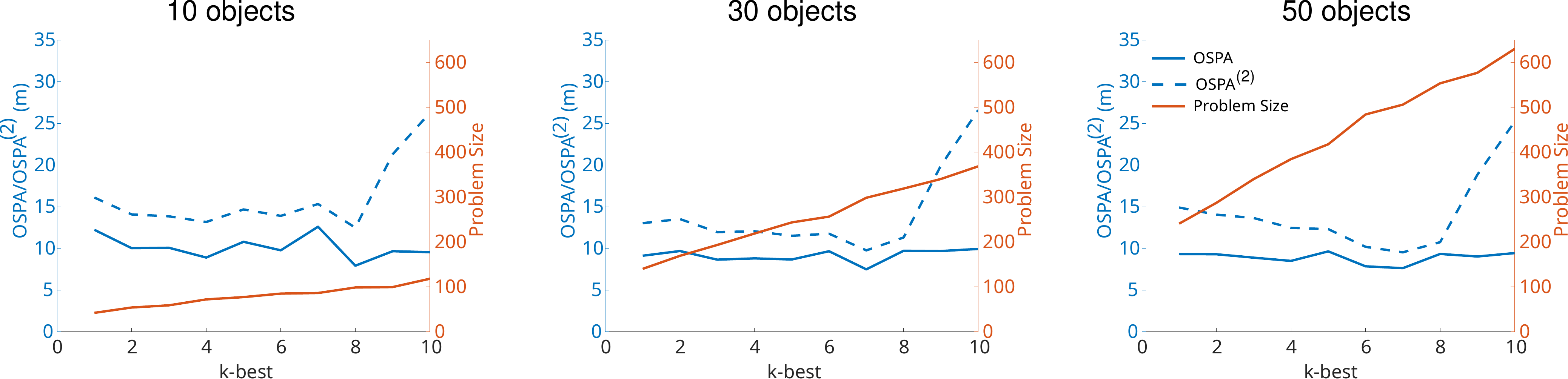}
	\caption{$k_{\text{best}}$ scaling with $10$, $30$ and $50$ objects.}
	\label{fig:k_scaling_exp}
\end{figure}

Figure~\ref{fig:k_scaling_exp} presents the average OSPA, OSPA\textsuperscript{(2)}, and the resulting assignment problem size measured as the average number of variables/hypotheses. Increasing $k_{\text{best}}$ leads to a noticeable reduction in OSPA\textsuperscript{(2)} while OSPA remains relatively stable. This behaviour suggests improved label consistency--- a reduction in identity switching ---in the estimated tracks, likely due to additional feasible hypotheses that help resolve close-proximity object interactions.

Nevertheless, as $k_{\text{best}}$ continues to increase, under all enumerations of problem size with increasing objects, the OSPA\textsuperscript{(2)} metric worsens. Although the observations appear counterintuitive, the effect is consistent with the discussion in Section~\ref{sec:clique}. For closely spaced objects, a larger $k_{\text{best}}$ can produce an excessive number of hypotheses without a corresponding increase in the discriminative power of the cost function. This can introduce additional ambiguous or low-quality hypotheses, increasing the likelihood of suboptimal assignments and, consequently, more frequent label switching---label consistency. Further, from a practical perspective, increasing $k_{\text{best}}$ is undesirable in resource-constrained settings as it leads to a substantial growth in the assignment problem size and computational cost, with no added benefit. Considering the trade-off between estimation performance and computational complexity, we use $k_{\text{largest}} = 5$ in our experiments.

\subsection{Taxi Tracking} \label{sec:exp2}
This experiment assesses the ability of the proposed MDAP-based fusion framework to maintain consistent object associations in real-world traffic monitoring scenarios where sensors have partial and overlapping fields of view. To this end, we evaluate the proposed distributed fusion algorithm using the CRAWDAD taxi dataset\footnote{\url{https://crawdad.org/epfl/mobility/20090224}} and the Porto taxi dataset\footnote{\url{https://kaggle.com/competitions/pkdd-15-predict-taxi-service-trajectory-i}}. The CRAWDAD taxi dataset contains taxi trajectories in the San Francisco Bay Area, USA, and the Porto taxi dataset contains trajectories in Porto, Portugal, released as part of the ECML/PKDD~2015 challenge. We use $20$ taxi tracks from each dataset and, following a similar approach to~\cite{dames_tracking,hoa2024multi}, compress the ground truth time scale by a factor of $10$, so that trajectories originally spanning over $\SI{1000}{\second}$ are tracked over approximately $\SI{100}{\second}$.

\begin{figure}[!tb]
	\centering \includegraphics[width=1.0\textwidth]{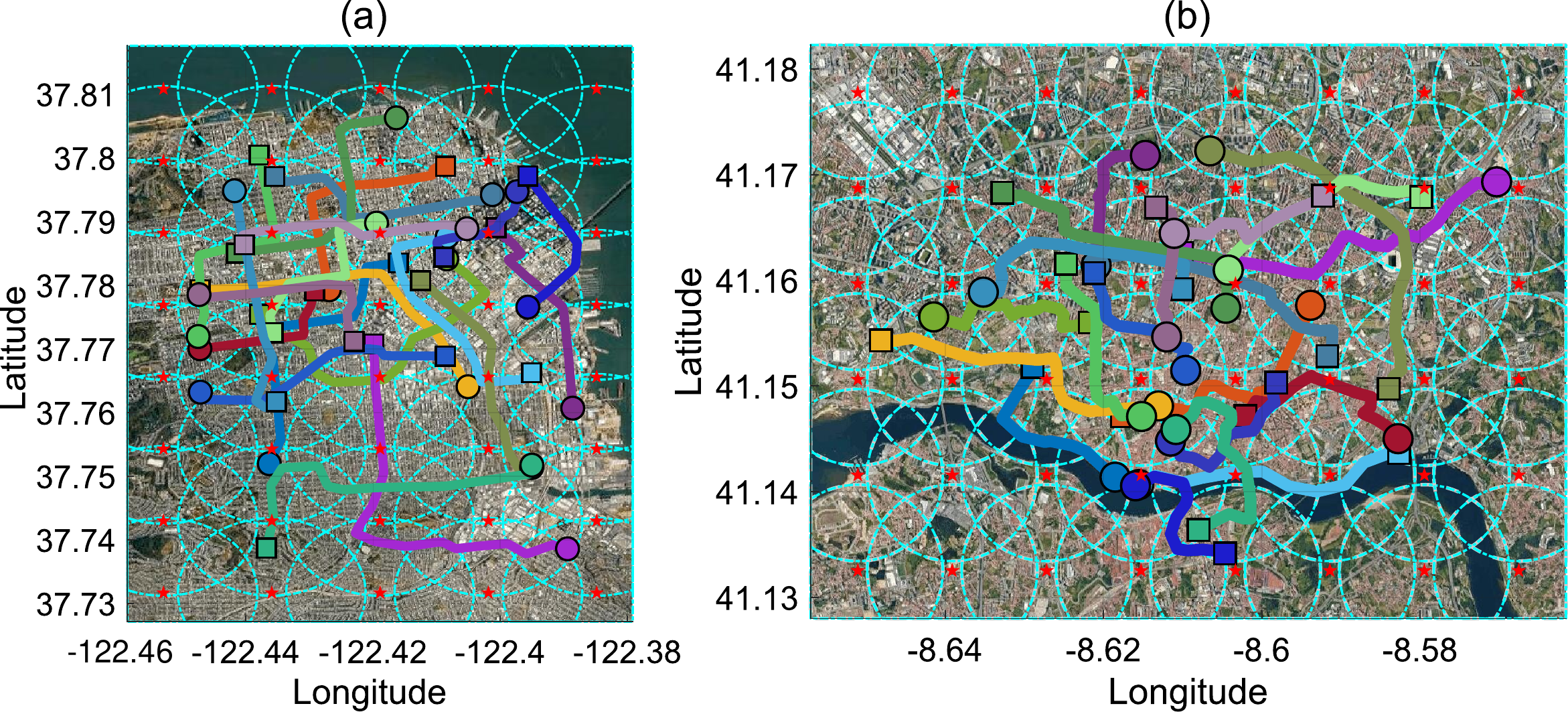}
	\caption{(a) San Francisco taxi tracking scenario.  The ground truth trajectories of $20$ taxis tracked by $40$ sensor nodes.  (b) Porto taxi scenario. Ground-truth trajectories of $20$ taxis tracked by $48$ sensor nodes. Starting and stopping positions are denoted by $\ocircle$ and $\Box$, respectively.}
	\label{fig:taxi_truth} 
\end{figure}

For the environment, we simulate a network of sensors with a limited FoV, a detection probability $P_{D} = 0.9$, and a clutter rate $\lambda_{c} = 3$ uniformly placed over the surveillance area. The San Francisco scenario spans $\SI{7000}{\meter}\times\SI{10000}{\meter}$ over a tracking period of $\SI{120}{\second}$, with $40$ sensor nodes of maximum detection range $\SI{1300}{\meter}$. The Porto taxi scenario spans $\SI{8000}{\meter}\times\SI{6000}{\meter}$ over $\SI{115}{\second}$, with $48$ sensor nodes of maximum detection range $\SI{900}{\meter}$. All sensor nodes are assumed to be network-connected, and $\gamma_{\max}=\SI{150}{\meter}$ is used for our proposed method. The ground truth tracks and sensor node positions for both scenarios are shown in Figure~\ref{fig:taxi_truth}.

\vspace{0.2cm}
\noindent\textbf{Object Dynamic Model.~}To approximate the behaviour of taxis, we employ the constant turn (CT) model with an unknown turn rate as the dynamic model. This model is defined as follows: 
\begin{align} \label{eq:ct_model}
	\begin{split}
		p_{k} &= F^{CT}(\theta_{k-1})p_{k-1} + G^{CT}\omega_{k} \\
		\theta_{k} &= \theta_{k-1} + T_{0}q_{k} \\
	\end{split}
\end{align}
where $x_{k}$ is the object state at time $k$ with $x = [p; \theta]$, $p = [\mathrm{x}, \dot{\mathrm{x}}, \mathrm{y}, \dot{\mathrm{y}}]^\top$, consisting of its 2D position, velocity and turning rate, $T_{0} = \SI{1}{\second}$, $\omega_{k}\sim \mathcal{N}(0_{2\times1}, \sigma_{\omega}^{2}I_{2})$ with $\sigma_{\omega} = \SI{5}{\meter/\second^{2}}$, $q_{k}\sim\mathcal{N}(0, \sigma_{q}^{2})$, and $\sigma_{q} = (\pi/8)\SI{}{\radian/\second}$,
\begin{align}
	F^{CT}(\theta) = \begin{bmatrix}
		1  &  \frac{\sin(\theta T_{0})}{\theta} & 0 & -\frac{1 - \cos(\theta T_{0})}{\theta} \\
		0  &  \cos(\theta T_{0})  &  0  & -\sin(\theta T_{0})  \\
		0  & \frac{1-\cos(\theta T_{0})}{\theta}  & 1  &  \frac{\sin(\theta T_{0})}{\theta}  \\
		0  & \sin(\theta T_{0})  &  0  &  \cos(\theta T_{0}) \\
	\end{bmatrix},\quad 
	G^{CT} = \begin{bmatrix}
		\frac{T_{0}^{2}}{2}  & 0 \\
		T_{0}  & 0 \\
		0  &  \frac{T_{0}^{2}}{2} \\
		0  &  T_{0} \\
	\end{bmatrix}.
\end{align}     
\noindent\textbf{Measurement Model.~}Each local node models a 2D-position sensor, such as a camera, \ie
\begin{align}
	z_k = H^{(\text{pos})}x_k + \eta^{(\text{pos})}_k,
\end{align}
where
\begin{align}
	H^{(\text{pos})} = 
	\begin{bmatrix}
		1 & 0 & 0 & 0 & 0 \\
		0 & 0 & 1 & 0 & 0 \\
	\end{bmatrix}
\end{align}
and $\eta^{(\textbf{pos})}_k$ is a $2\times1$ zero mean Gaussian measurement noise, \ie, $\eta^{(\textbf{pos})}_k \sim \mathcal{N}(0_{2\times1}, (\sigma^{(\text{pos})}_\eta)^2 I_2)$ with $\sigma^{(\text{pos})}_\eta = 10~\text{m}.$

\vspace{2mm}
\noindent\textbf{Experiment Results.}

\begin{table}[!b]
\centering
\setlength{\extrarowheight}{0pt}
\addtolength{\extrarowheight}{\aboverulesep}
\addtolength{\extrarowheight}{\belowrulesep}
\setlength{\aboverulesep}{0pt}
\setlength{\belowrulesep}{0pt}
\caption{San Francisco and Porto taxi tracking scenario results summary. Results are reported as mean $\pm$ one standard deviation over 100 Monte Carlo trials for each method. Bold text indicates the best result.}
\label{tab:taxi_table}
\resizebox{\linewidth}{!}{
\begin{tabular}{lccc|ccc} 
\toprule
\rowcolor[rgb]{0.89,0.89,0.89} \multicolumn{4}{c}{\textbf{San Francisco Taxi Tracking Scenario}} & \multicolumn{3}{c}{\textbf{Porto Taxi Tracking Scenario}}                        \\ 
\midrule
              & OSPA (m)                & OSPA\textsuperscript{(2)} Dist (m) & Fusing time (ms)  & OSPA (m)                & OSPA\textsuperscript{(2)} Dist (m) & Fusing time (ms)  \\ 
\midrule
\cdp{}        & $27.02\pm1.12$          & $54.49\pm2.00$                     & $79\pm17$         & $26.10\pm0.82$          & $49.17\pm1.54$                     & $63\pm14$         \\
\meanshift{}  & $28.19\pm0.94$          & $56.15\pm2.32$                     & $79\pm16$         & $27.61\pm0.82$          & $50.66\pm2.22$                     & $63\pm13$         \\
\our{} (Ours) & $\mathbf{26.75\pm0.94}$ & $\mathbf{49.61\pm2.05}$            & $87\pm17$         & $\mathbf{25.88\pm0.97}$ & $\mathbf{44.59\pm1.82}$            & $72\pm15$         \\ 
\hline
\multicolumn{7}{l}{{\cellcolor[rgb]{1,1,1}}\textit{Centralised reference}}\\ 
\hline
MS-LMB        & $21.43\pm1.48$          & $28.82\pm1.69$                     & N/A               & $21.21\pm0.92$          & $28.76\pm1.51$                     & N/A               \\
\bottomrule
\end{tabular}
}
\end{table}

\begin{figure}[!tb]
	\centering \includegraphics[width=1.0\textwidth]{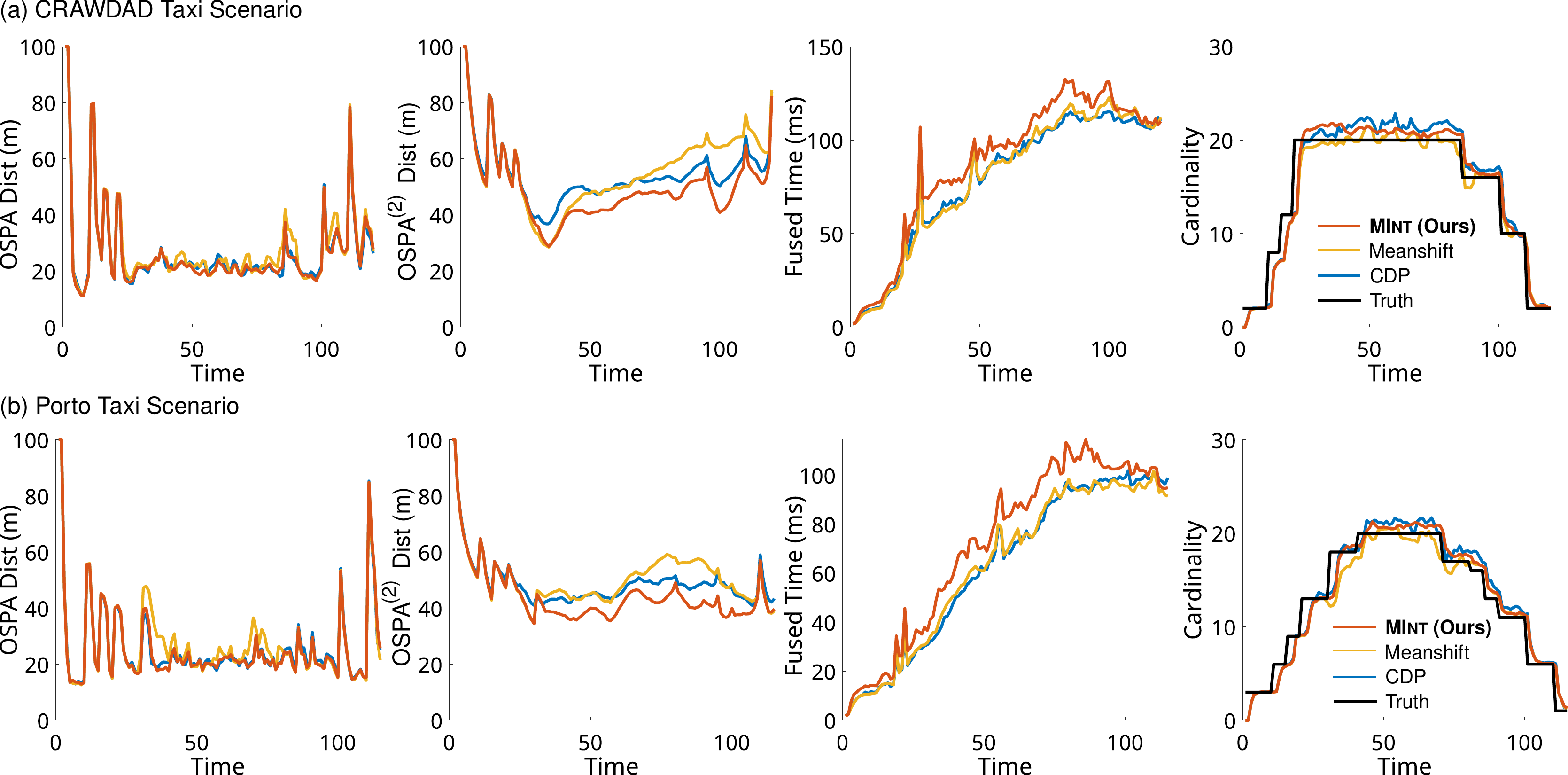}
	\caption{San Francisco and Porto taxi tracking scenario. Averaged OSPA, OSPA\textsuperscript{(2)} Distance, Cardinality and fusing time over $100$ MC runs. Each node employs an LMB filter for (a) San Francisco taxi tracking scenario; (b) Porto taxi tracking scenario.}
	\label{fig:taxi_stat} 
\end{figure}

Table~\ref{tab:taxi_table} and Figure~\ref{fig:taxi_stat} present detailed performance comparisons across all MC executions of both taxi tracking scenarios.
Overall, despite the challenging dynamic behaviour of taxis, our \our{} algorithm performs better than \cdp{} and the alternative cluster analysis with the \meanshift{} method in terms of estimation and tracking accuracy, as shown in the OSPA and OSPA\textsuperscript{(2)} results, respectively, in Table~\ref{tab:taxi_table}.
The margin is, however, markedly different for the two metrics. Relative to \cdp{}, the OSPA improvement is approximately $1\%$ in both scenarios, whereas the OSPA\textsuperscript{(2)} improvement is around $10\%$ in both scenarios, rising to $13\%$ when compared against \meanshift{}. Since OSPA is evaluated per scan while OSPA\textsuperscript{(2)} is evaluated over a sliding window of the estimated tracks, this pattern indicates that all three methods localise the taxis with comparable per-scan accuracy and that the performance gap arises almost entirely from the consistency of the fused labels over time. 

The cardinality estimates in Figure~\ref{fig:taxi_stat} identify the source of this inconsistency.
\cdp{} identifies valid associations by employing a k-means (k=2) analysis and \meanshift{} by kernel bandwidth, where the number of fused objects is re-estimated at every step, and neither constrains a fused object to at most one track per node. Consequently, both are poorly conditioned here. First, convoying taxis on a road network forms elongated, unequal groups rather than the compact clusters favoured by cluster analysis methods. 
%k-means favours. 
An overestimated count splits a single taxi's tracks between two clusters, creating a spurious object, consistent with the observed cardinality overestimation. Such objects impact the OSPA metric only through its cardinality term, but the repeated re-estimation destabilises the fused labels across scans, and the accumulated switches explain the much larger OSPA\textsuperscript{(2)} gap. In contrast, the proposed formulation does not rely on a cluster-analysis count, explicitly constrains the problem by imposing a one-track-per-node limit, and yields an optimal assignment over the retained candidate hypotheses. Consequently, a fused label persists even as its supporting nodes change, and its cardinality remains stable. Importantly, we observe the label persistence to hold across both scenarios; notably, the results are achieved without retuning or reliance on scenario-specific parameters for the algorithm.

\begin{figure}[!tb]
	\centering \includegraphics[width=0.95\textwidth]{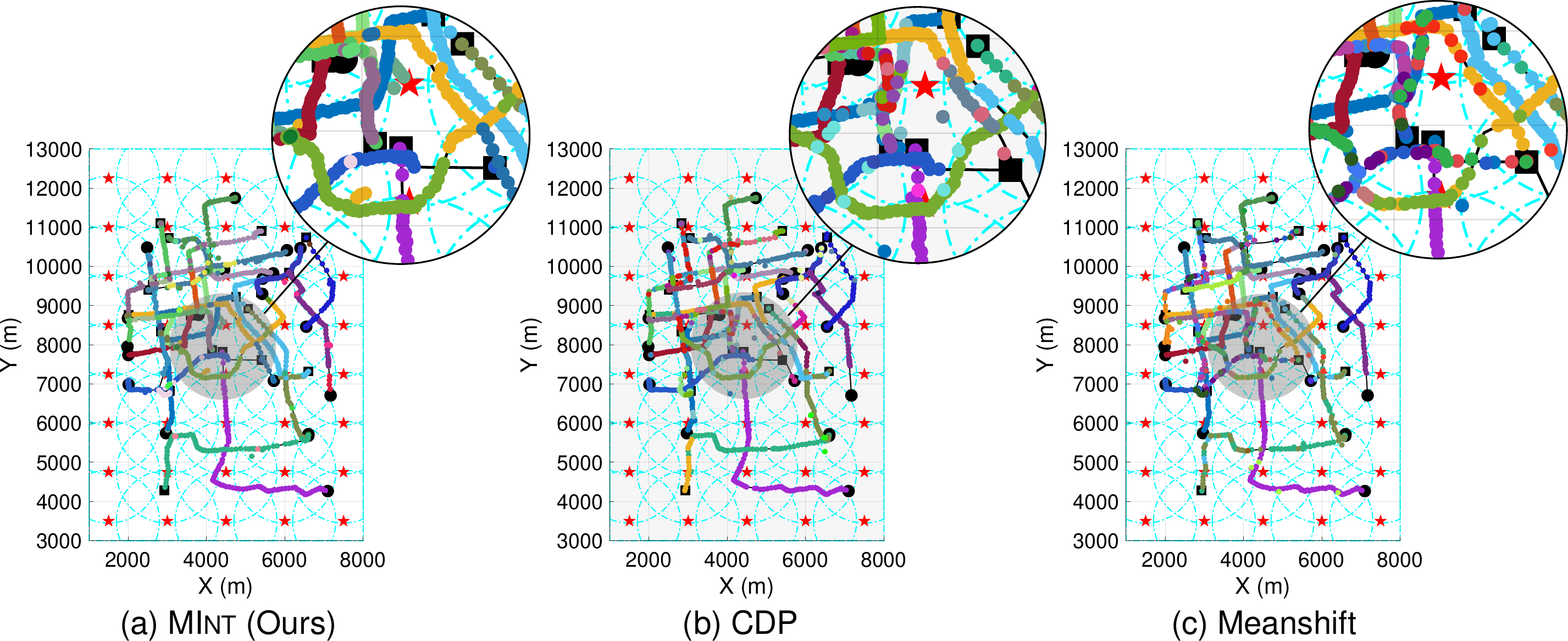}
	\caption{San Francisco taxi tracking scenario. An instance of estimated trajectories using \our{}, \cdp{} and \meanshift{} algorithms with a local LMB filter.}
	\label{fig:taxi_est_sample} 
\end{figure}

\begin{figure}[!tb]
	\centering \includegraphics[width=1.0\textwidth]{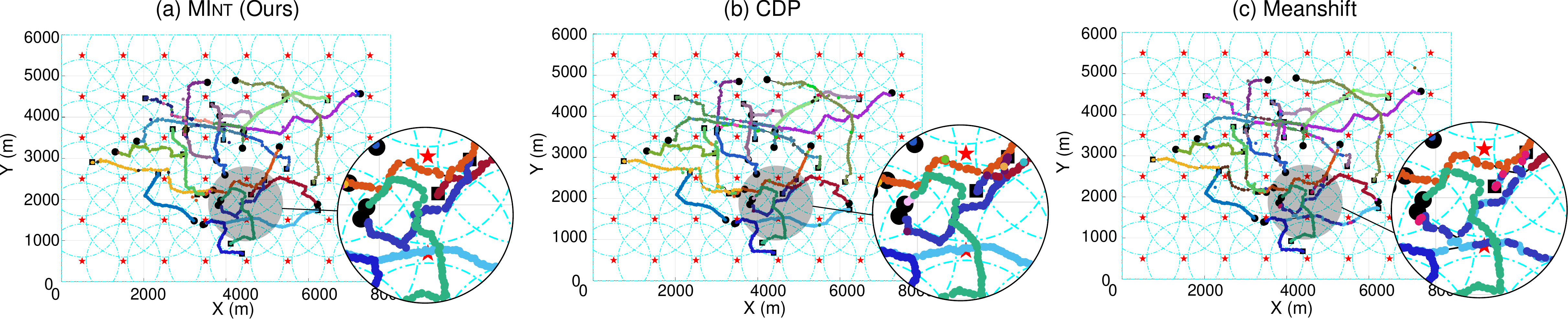}
	\caption{Porto taxi tracking scenario. An instance of estimated trajectories using \our{}, \cdp{} and \meanshift{} algorithms with a local LMB filter.}
	\label{fig:ECML_est_sample} 
\end{figure}

Figure~\ref{fig:taxi_est_sample} and Figure~\ref{fig:ECML_est_sample} illustrate this behaviour for a tracking experiment, depicting the labelled multi-object state estimates at node $1$. Each track colour denotes a distinct fused label. All three methods recover the taxis' trajectories, and the enlarged regions in the insets, taken at locations where several taxis traverse the same junction, isolate differences in label consistency. Under \our{}, each trajectory retains a single colour throughout the enlarged region; in contrast, under \cdp{} and \meanshift{} several trajectories change colour part-way along, indicating that the fused label has been reassigned mid-track. The effect is most pronounced for \meanshift{}, and this observation is consistent with its larger OSPA\textsuperscript{(2)} in Table~\ref{tab:taxi_table}.

Importantly, the fusing time results in Table~\ref{tab:taxi_table} demonstrate that the exact assignment is obtained at small cost, with \our{} adding \SI{8}{\milli\second} per fusion step in the San Francisco scenario and \SI{9}{\milli\second}  in the Porto scenario relative to \cdp{}. The overhead remains bounded because gating restricts the candidate track tuples to spatially proximate tracks, so the assignment decomposes into small independent sub-problems rather than scaling with the total number of fused tracks. The fusing times of all three methods are also lower in the Porto scenario, which follows from its shorter detection range and the correspondingly smaller regions with mutually overlapping FoVs.

\subsection{Drone Tracking Scenario in 3D}\label{sec:exp-drones}
\begin{figure}[!tb]
	\centering
	\includegraphics[width=1.0\textwidth]{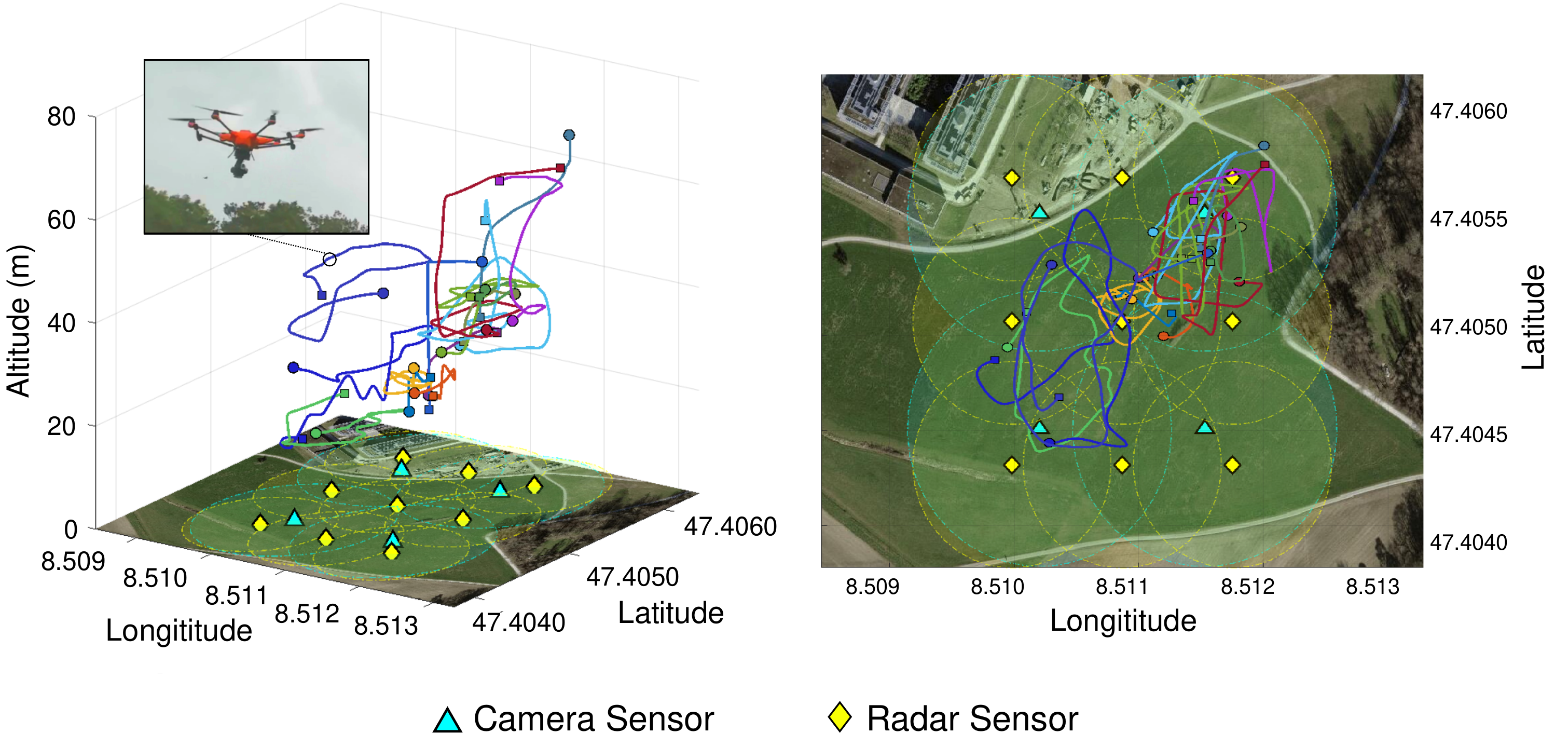}
	\caption{Drone Tracking~\cite{jingtong2020} ground truth and sensor placement.}
	\label{fig:dronetracking_truth}
\end{figure}

As a more challenging problem, we consider a drone tracking scenario in a three-dimensional space with heterogeneous sensor coverage. The setting evaluates the robustness of the proposed framework under more complex spatial conditions and demonstrates its ability to fuse track estimates from sensors with different coverage and sensor characteristics. The drone tracking dataset reported in~\cite{jingtong2020} is employed to evaluate the proposed algorithm under realistic three-dimensional operating conditions. The dataset comprises high-precision drone trajectories within a spatial region of $\SI{200}{\meter} \times \SI{250}{\meter} \times \SI{70}{\meter}$. A total of $15$ drone trajectories are used for the experiments with $\gamma_{\max}=\SI{16}{\meter}$ along with $9$ simulated radar sensors and $4$ camera sensors deployed within the tracking zone. All of the sensor nodes are assumed to be network-connected. The ground-truth drone trajectories and sensor deployments are illustrated in Figure~\ref{fig:dronetracking_truth}. 

\vspace{2mm}
\noindent\textbf{Object Dynamic Model.~} To model the behaviour of the drone trajectory, a constant velocity model is used to describe the drone's trajectory as follows:
\begin{align}\label{eq:ct3_model}
	\begin{split}
		x_{k} &= F^{(CV)}x_{k-1} + \omega_{k} \\
        F^{(CV)} &= I_{3} \otimes \begin{bmatrix}
            1 & T_{0} \\
            0 & 1
        \end{bmatrix}  
	\end{split}
\end{align}
where $x_{k}$ is the object state at time $k$ with $x = [\mathrm{x}, \dot{\mathrm{x}}, \mathrm{y}, \dot{\mathrm{y}}, \mathrm{z}, \dot{\mathrm{z}}]^\top$, consisting of its 3D position and velocity,  $\otimes$ denotes the Kronecker tensor product operator between two matrices, $T_{0} = \SI{1}{\second}$, $\omega_{k}\sim \mathcal{N}(0_{6\times1}, Q^{CV})$ is a $6\times1$ zero mean Gaussian process noise with covariance and $Q^{CV} = I_{3}\otimes\sigma_{CV}^{2}\begin{bmatrix}T_{0}^{4}/4 & T_{0}^{3}/2 \\ T_{0}^{3}/2 & T_{0}^{2}\end{bmatrix} $, with $\sigma_{CV}=\SI{0.35}{\meter/\second^{2}}$.

\vspace{2mm}
\noindent\textbf{Measurement Model.~}
The radar sensors in this scenario produce 3D radar measurements, including range, elevation angle, and azimuth, \ie
\begin{align}
	z_k = h^{(\text{radar})}(x_k) + \eta^{(\text{radar})}_k
\end{align}
where 
\begin{align}
	h^{(\text{radar})}(x)= \begin{bmatrix}
		\sqrt{\mathrm{x}^2 + \mathrm{y}^2 + z^{2}} \\
		\mathrm{atan2}\big(z,\sqrt{x^{2}+y^{2}} \big) \\
		\mathrm{atan2}\big(\mathrm{y},\mathrm{x} \big)
	\end{bmatrix}
\end{align}
and $\eta^{(\text{radar})}_k$ is a $3\times1$ zero mean Gaussian measurement noise,~\ie, $\eta^{(\text{radar})}_k \sim \mathcal{N}(0_{3\times1}, \operatorname{diag}(\sigma^{(\text{radar})}_\eta)^{2})$, with $\sigma^{(\text{radar})}_\eta = [1~\text{m},\pi/360~\text{rad},\pi/360~\text{rad}]^{\top}$, and $\mathrm{atan2}(\cdot, \cdot)$ being the four-quadrant inverse tangent function.

Radar sensors are setup to have a detection probability $P_{D}^{\text{(radar)}} = 0.9$, clutter rate $\lambda_{c}^{\text{(radar)}}=3$ and maximum detection range of $\SI{60}{\meter}$.

The camera sensors are modelled as a 3D-position sensor with maximum detection range of $\SI{80}{\meter}$, \ie{}
\begin{equation}
	z_{k} = H^{\text{(pos)}}x_{k} + \eta_{k}^{\text{(pos)}}
\end{equation}
where $H^{\text{(pos)}} = I_{3}\otimes[1,0]$, $\eta^{(\text{pos})}_k \sim \mathcal{N}(0_{3\times1}, {\sigma^{(\text{pos})}_\eta}^{2}I_{3})$, with $\sigma^{(\text{pos})}_\eta = 1~\text{m}$. Camera sensors' detection probability $P_{D}^{\text{(camera)}} = 0.8$ and clutter rate $\lambda_{c}^{\text{(camera)}} = 1$.

\begin{table}
\centering
\setlength{\extrarowheight}{0pt}
\addtolength{\extrarowheight}{\aboverulesep}
\addtolength{\extrarowheight}{\belowrulesep}
\setlength{\aboverulesep}{0pt}
\setlength{\belowrulesep}{0pt}
\caption{3D Drone tracking scenario results summary. Results are reported as mean $\pm$ one standard deviation over 100 Monte Carlo trials for each method. Bold text indicates the best result.}
\label{tab:drone_table}
\begin{tabular}{lccc} 
\toprule
\multicolumn{4}{c}{{\cellcolor[rgb]{0.89,0.89,0.89}}\textbf{3D Drone Tracking Scenario}}         \\ 
\midrule
              & OSPA (m)                & OSPA\textsuperscript{(2)} Dist (m) & Fusing time (ms)  \\ 
\midrule
\cdp{}        & $13.55\pm0.58$          & $32.55\pm2.56$                     & $65\pm14$         \\
\meanshift{}  & $15.42\pm0.54$          & $36.92\pm2.35$                     & $65\pm14$         \\
\our{} (Ours) & $\mathbf{9.88\pm0.54}$ & $\mathbf{26.50\pm2.42}$            & $72\pm14$         \\ 
\hline
\multicolumn{4}{l}{{\cellcolor[rgb]{1,1,1}}\textit{Centralised reference}}              \\ 
\hline
MS-LMB        & $9.30\pm0.83$          & $14.18\pm1.39$                     & N/A               \\
\bottomrule
\end{tabular}
\end{table}

\begin{figure}[!tb]
	\centering
	\includegraphics[width=1.0\textwidth]{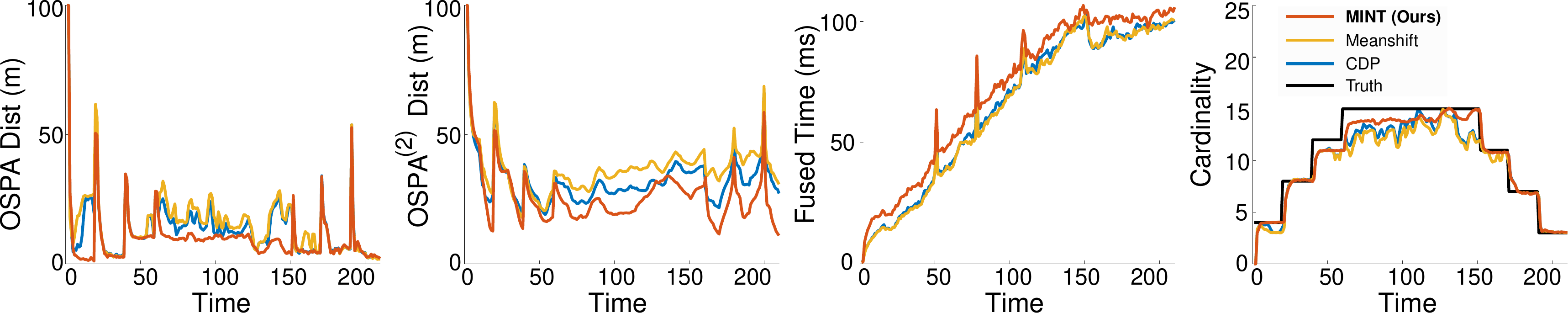}
	\caption{3D Drone tracking scenario. OSPA, OSPA\textsuperscript{(2)}, fusing time and cardinality results averaged over $100$ Monte Carlo trials.}
	\label{fig:dronetracking_result}
\end{figure}

\begin{figure}[!tb]
	\centering
	\includegraphics[width=1.0\linewidth]{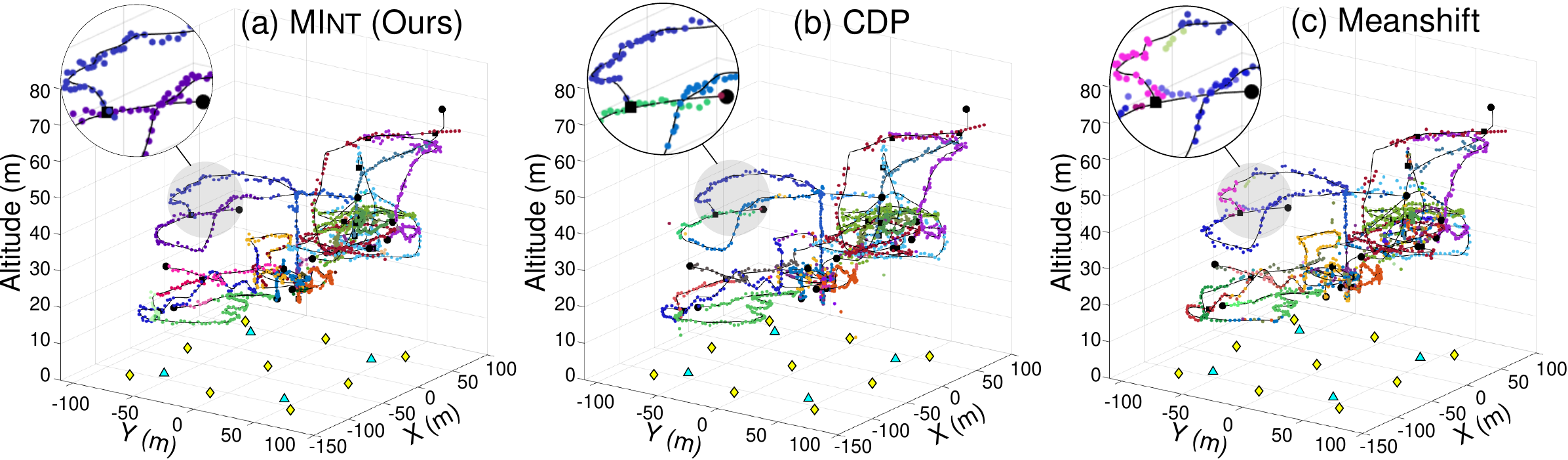}
	\caption{3D Drone tracking scenario. An instance of estimated trajectories from a single Monte Carlo execution, comparing our proposed \our{} algorithm with the \cdp{} and \meanshift{} based fusion algorithms.}
	\label{fig:dronetracking_est}
\end{figure}

\vspace{2mm}
\noindent\textbf{Experiment Results.~}

Table~\ref{tab:drone_table} and Figure~\ref{fig:dronetracking_result} present the Monte Carlo simulation results, while Figure~\ref{fig:dronetracking_est} shows a fusion example from a single execution. Consistent with the results reported in the preceding experimental sections, the proposed \our{} algorithm achieves superior tracking performance across both the OSPA and OSPA\textsuperscript{(2)} metrics.
The cardinality plot in Figure~\ref{fig:dronetracking_result} accounts for this ordering. Only \our{} tracks the true staircase closely and over the plateau of $15$ objects from 50~s to 150~s; \meanshift{} settles at several objects below the truth throughout, and \cdp{} by a smaller margin. With $15$ drones manoeuvring within a volume of roughly $\SI{200}{\meter}\times\SI{250}{\meter}\times\SI{70}{\meter}$, object separation frequently falls below the kernel bandwidth, so \meanshift{} analysis merges distinct drones into a single mode. Each merged pair costs manifests as a missed object in the per-scan OSPA metric and, as the surviving track inherits only one label, as a label inconsistency that persists across the OSPA\textsuperscript{(2)} window to result in the larger gap in the latter metric.

The observed underestimation in cardinality for cluster-based analysis methods---missed objects---is a reversal of the overestimation seen in the taxi scenarios in Section~\ref{sec:exp2}, Figure~\ref{fig:taxi_stat}. A single fixed spatial scaling used in cluster analysis fails to account for opposing effects according to object density. \our{} avoids this sensitivity, as the number of fused objects follows from the assignment itself, with candidate tuples admitted for candidate hypothesis generation and the error bound of Proposition~\ref{prop:1}, rather than by a tuned bandwidth or cluster-analysis count. Figure~\ref{fig:dronetracking_est} demonstrates the contrasting effect of the methods. In the enlarged inset regions, where two drone flights lay in close proximity, \our{} estimates persist a single label for each trajectory. In contrast, the \cdp{} and \meanshift{} estimates switch labels, indicated by a change in colours part-way along. The fusing times remain nearly coincident throughout, so the exact assignment formulation we pursued does not become disproportionately expensive as cardinality increases.

\section{Conclusions} \label{sec:conclusion}
In this article, we formulate the fusion problem in distributed multi-object tracking with unknown and limited-field-of-view sensors, pertinent to practical settings, as a multi-dimensional assignment problem.
We systematically develop a theoretical framework, transforming the problem from its intractable form to a formulation solvable using integer linear programming. We introduce a cost formulation that relates fusion costs to estimation errors, enabling computationally tractable global optimisation and principled selection of gating parameters. We introduce a systematic hypothesis-generation strategy based on geometric sparsity and clique discovery to reduce the MDAP search space prior to facilitating optimisation using an ILP formulation. Overall, the method enables efficient globally consistent association decisions across distributed sensor nodes.

Extensive comparative evaluations on real-world taxi and drone datasets, tracking problems in both two- and three-dimensional spaces involving multiple sensors with partial and overlapping coverage, demonstrate that the proposed approach achieves improved tracking accuracy compared with state-of-the-art DMOT methods while maintaining comparable computational efficiency.  

Despite the advances made in the study towards addressing a challenging problem, comparisons with the centralised reference indicate a notable relative performance margin with the OSPA$^{(2)}$ metric. While a performance difference with a centralised reference, representative of an idealised baseline, is expected, the difference observed in OSPA$^{(2)}$ compared to OSPA, is suggestive of the margin for improvement possible with improving label consensus. 
Label consensus is a challenging problem by itself and the approach we adopted from prior work may limit track-label consistency in complex multi-object scenarios. Improving label consistency through label management methods remains an important direction for future research. 

\section{Acknowledgement}
This work was partially supported by grant funding from The Defence Science and Technology Group (DSTG) through the Centre for Advanced Defence Research in Robotics and Autonomous Systems (CADR-RAS), the Office of Naval Research Global (N629092512046) and the Australian Research Council (ARC) (LP200301507).%  [XXXXX]

\addcontentsline{toc}{section}{References}  \bibliographystyle{IEEEtran}
\bibliography{references}

\end{document}